%% file: ms.tex
\documentclass[twocolumn]{aastex6}

\let\pwiflocal=\iffalse \let\pwifjournal=\iffalse
\input{setup}
\newcommand{\name}{K2-33}
\newcommand{\pname}{K2-33b}
\newcommand{\prad}{$5.04^{+0.34}_{-0.37}$\,R$_\earth$}
\newcommand{\ktwo}{{\it K2}}
\newcommand{\mearth}{MEarth}
\newcommand{\av}{$A_V$}

\newcommand\ut{1}
\newcommand\hub{2}
\newcommand\har{3}
\newcommand\nsf{4}
\newcommand\up{5}
\newcommand\geo{6}
\newcommand\ctio{7}
\newcommand\ifa{8}
\newcommand\wash{9}
\newcommand\anu{10}

\slugcomment{Accepted to AJ}

\shorttitle{A Transiting Planet in Upper Scorpius}

\shortauthors{Mann et al.}

\begin{document}
 
\title{Zodiacal Exoplanets in Time (ZEIT) III:\\ A short-period planet orbiting a pre-main-sequence star in the Upper Scorpius OB Association}

\author{Andrew W. Mann\footnote{\email{amann@astro.as.utexas.edu}}\altaffilmark{\ut, \hub}, Elisabeth R. Newton\altaffilmark{\har,\nsf}, Aaron C. Rizzuto\altaffilmark{\ut}, Jonathan Irwin\altaffilmark{\har}, Gregory A. Feiden\altaffilmark{\up}, Eric Gaidos\altaffilmark{\geo}, Gregory N. Mace\altaffilmark{\ut}, Adam L. Kraus\altaffilmark{\ut}, David J. James\altaffilmark{\ctio}, Megan Ansdell\altaffilmark{\ifa}, David Charbonneau\altaffilmark{\har}, Kevin R. Covey\altaffilmark{\wash}, Michael J. Ireland\altaffilmark{\anu}, Daniel T. Jaffe\altaffilmark{\ut}, Marshall C. Johnson\altaffilmark{\ut}, Benjamin Kidder\altaffilmark{\ut}, Andrew Vanderburg\altaffilmark{\har,\nsf}}

\altaffiltext{\ut}{Department of Astronomy, The University of Texas at Austin, Austin, TX 78712, USA}
\altaffiltext{\hub}{Hubble Fellow}
\altaffiltext{\har}{Harvard-Smithsonian Center for Astrophysics, Cambridge, Massachusetts 02138, USA}
\altaffiltext{\nsf}{NSF Graduate Research Fellow}
\altaffiltext{\up}{Department of Physics and Astronomy, Uppsala University, Uppsala, Sweden}
\altaffiltext{\geo}{Department of Geology \& Geophysics, University of Hawaii at Manoa, Honolulu, HI 96822, USA}
\altaffiltext{\ctio}{Cerro Tololo Inter-American Observatory, Colina el Pino, LaSerena, Chile}
\altaffiltext{\ifa}{Institute for Astronomy, University of Hawaii at Manoa, Honolulu, HI 96822, USA}
\altaffiltext{\wash}{Western Washington University, Department of Physics \& Astronomy, Bellingham, WA 98225, USA}
\altaffiltext{\anu}{Research School for Astronomy \& Astrophysics, Australia National University, Canberra ACT 2611, Australia}

\begin{abstract}
We confirm and characterize a close-in ($P_{\rm{orb}}$ = 5.425 days), super-Neptune sized ($5.04^{+0.34}_{-0.37}$\,R$_\earth$) planet transiting \name\ (2MASS~J16101473-1919095), a late-type (M3) pre-main sequence (11 Myr-old) star in the Upper Scorpius subgroup of the Scorpius-Centaurus OB association. The host star has the kinematics of a member of the Upper Scorpius OB association, and its spectrum contains lithium absorption, an unambiguous sign of youth ($<20$~Myr) in late-type dwarfs. We combine photometry from K2 and the ground-based MEarth project to refine the planet's properties and constrain the host star's density. We determine \name's bolometric flux and effective temperature from moderate resolution spectra. By utilizing isochrones that include the effects of magnetic fields, we derive a precise radius (6-7\%) and mass (16\%) for the host star, and a stellar age consistent with the established value for Upper Scorpius. Follow-up high-resolution imaging and Doppler spectroscopy confirm that the transiting object is not a stellar companion or a background eclipsing binary blended with the target. The shape of the transit, the constancy of the transit depth and periodicity over 1.5 years, and the independence with wavelength rules out stellar variability, or a dust cloud or debris disk partially occulting the star as the source of the signal; we conclude it must instead be planetary in origin. The existence of \pname\ suggests close-in planets can form in situ or migrate within $\sim 10$~Myr, e.g., via interactions with a disk, and that long-timescale dynamical migration such as by Lidov-Kozai or planet-planet scattering is not responsible for all short-period planets. 
\end{abstract}


\maketitle

\section{Introduction}\label{sec:intro}
Many known exoplanets orbit within 0.1\,AU of their host star, where they are more readily detected by the transit and Doppler methods \citep[e.g.,][]{2010Sci...330..653H,Fressin:2013qy}. Whether these planets formed near their present position (in situ), i.e., from circumstellar material interior to 0.5 AU \citep[e.g.,][]{2013MNRAS.431.3444C, 2015A&A...578A..36O}, or accreted at distances $>1$\,AU, and later migrated inwards \citep[e.g.,][]{2009ApJ...691.1322S, 2014MNRAS.440L..11R} is actively debated. If the planets migrated, the physical mechanism(s) behind their migration is yet another further point of debate.

Mechanisms of planet migration can be loosely divided into three categories: interactions with the protoplanetary disk \citep[e.g.,][]{2008ApJ...673..487I, 2010arXiv1004.4137L}, interactions between a stellar companion and the planet \citep[the Lidov-Kozai mechanism, e.g.,][]{2003ApJ...589..605W}, and interactions among multiple planets \citep[e.g.,][]{2006ApJ...638L..45F, 2008ApJ...686..580C}. Disk migration must occur before the protoplanetary disk dissipates/photo-evaporates ($\lesssim$10\,Myr) \citep[e.g.,][]{1997ApJ...482L.211W}. Migration involving angular momentum exchange with a third body typically operates on timescales much longer than disk migration ($\simeq$100\,Myr to more than 1\,Gyr), depending on the orbital and physical properties of the planet and perturber \citep{Fabrycky:2007ys, 2008ApJ...678..498N}. 

The difference in timescales presents a possible method to distinguish between these migration mechanisms. Close-in super-Earths or Jupiter-size planets around stars younger than 100\,Myr could not have migrated by a slow process like planet-planet or planet-star interaction, and instead likely formed in situ or migrated quickly through interaction with the disk. A comparison of the occurrence rate of close-in planets over a range of ages (10-1000\,Myr) would constrain the fraction of planets migrating on a given timescale.

High-precision photometry offers the best opportunity to detect the close-in planets needed to test migration theories \citep[e.g.,][]{1996JGR...10114853J}. Such planets are too close to their host star to be detected by direct imaging \citep[e.g.,][]{2012A&A...541A..89L,Bowler2015b}. Starspot-induced jitter complicates the detection of the planetary reflex motion \citep{2011ApJ...736..123M,2004AJ....127.3579P}, such that radial velocity (RV) surveys of young stars primarily uncover hot Jupiters around stars older than 100\,Myr \citep[e.g.,][]{Quinn:2012mz,Quinn2014}. Spot modulation can generate complicated variations in the light curve that makes detecting transiting planets more difficult. However, spots and transits create characteristically different patterns in a light curve which can be separated with precise photometry. Indeed the only close-in planet (candidates) around $<20$\,Myr old stars are from transit surveys \citep[e.g.,]{2012ApJ...755...42V,2012AJ....143...72M,2015MNRAS.446..411K}.

The repurposed \kepler\ mission \citep[\ktwo, ][]{Howell2014} has the photometric precision (tens of ppm) and observational baseline (70-80 days) required to detect small planets and rule out false-positive signals related to stellar youth (e.g., debris disks, and spots). We are carrying out a search for planets transiting stars in 10-800\,Myr young open clusters and OB associations using \ktwo. Our survey, Zodiacal Exoplanets in Time, includes Upper Scorpius \citep[$\simeq$11\,Myr, ][]{2012ApJ...746..154P, 2016ApJ...817..164R}, Taurus \citep[0-5\,Myr, ]{Kenyon2008}, Pleiades \citep[$\simeq$125\,Myr, ][]{2015ApJ...813..108D}, and Praesepe and Hyades \citep[650-800\,Myr, ][]{Brandt2015}. Our goal is to better understand if planets evolve from infancy (1-10\,Myr) to maturity ($\gtrsim$1\,Gyr), including how planets migrate, how, and on what timescales. 

Here we confirm and characterize a \prad\ planet (\pname) orbiting the pre-main-sequence (PMS) star \name\ (2MASS J16101473-1919095, EPIC~205117205), a member of the Upper Scorpius subgroup of the Scorpius-Centaurus (Sco-Cen) OB association. \pname\ was previously identified as a planet candidate by \citet{2016ApJS..222...14V}, but assigned inaccurate stellar and planetary parameters owing to the assumption of a main sequence age and an unreddened spectral energy distribution for the host star. In Section~\ref{sec:obs} we describe our follow-up observations, including moderate- to high-resolution spectroscopy, adaptive optics imaging and non-redundant aperture masking, and transit photometry. Our analysis of the light curve data is described in Section~\ref{sec:lc}. In Section~\ref{sec:params} we derive parameters for the host star. We use the available data to confirm a planetary origin of the transit signal, as we describe in Section~\ref{sec:fpp}. We conclude in Section~\ref{sec:discussion} with a brief summary and discussion of the importance of this planet.

\section{Observations and data reduction}\label{sec:obs} 

\subsection{\ktwo\ observations and light curve extraction}
From 2014 August 23 to 2014 November 13 (Campaign 2) \ktwo\ observed the core of Upper Scorpius, including \name. Owing to the loss of two reaction wheels the \kepler\ satellite drifts. To correct the pointing, \kepler's thrusters fire every $\sim$6\,hours. However, during the drift and subsequent thruster fire a stellar image usually moves over the detector. Combined with variations in the pixel sensitivity this generates changes in total measured flux from a given star as a function of centroid position. 

\begin{figure*}
	\centering
	\includegraphics[width=0.95\textwidth,trim={0 5cm 0 0},clip]{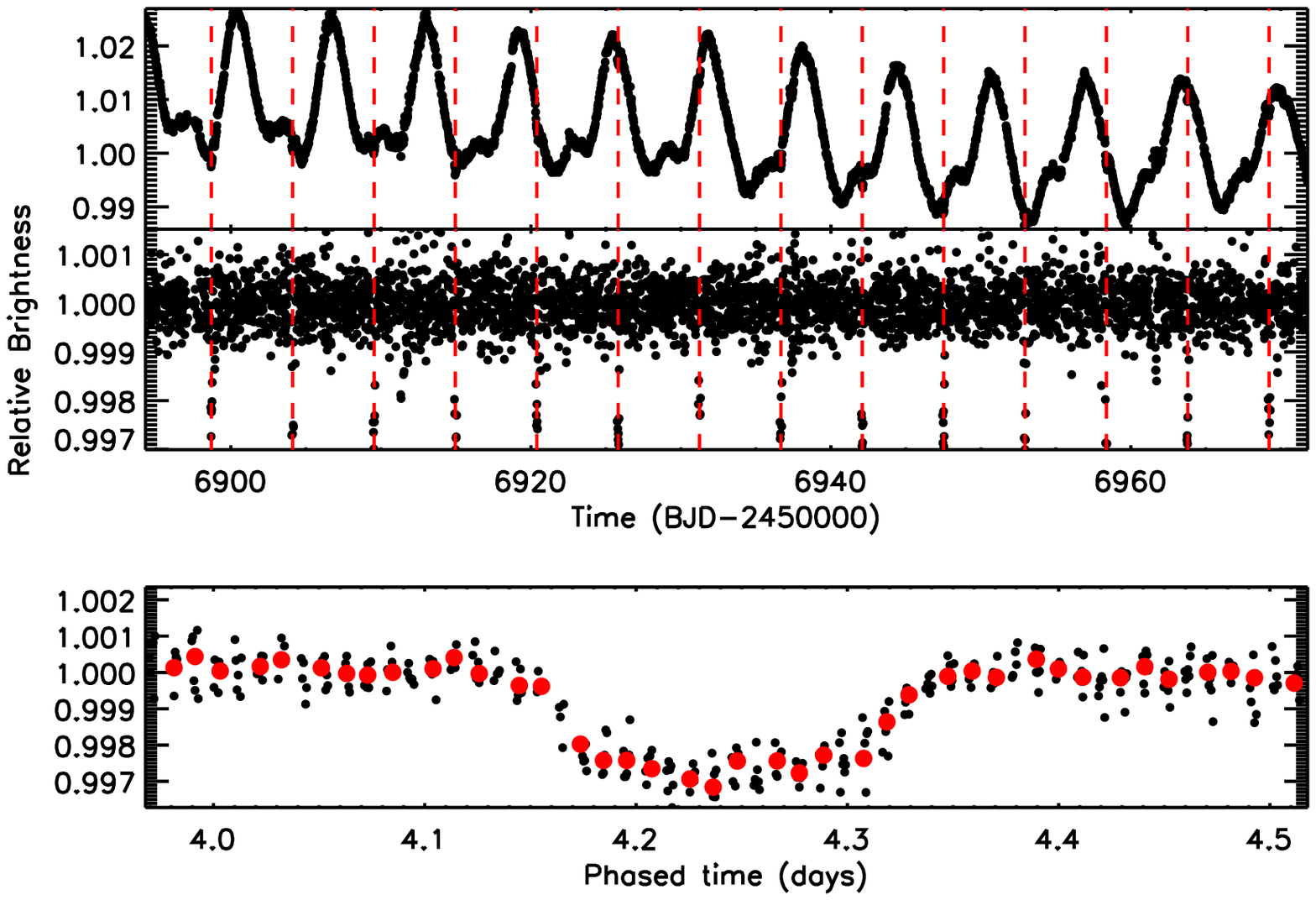} 
	\caption{Light curve of \name\ taken by the \kepler\ spacecraft. The top panel displays the light curve constructed from \kepler\ pixel data after removing effects of \kepler\ roll. The bottom panel shows the light curve after removing stellar variability. Red dashed lines indicate transits. Both curves are normalized to one.}
	\label{fig:lc}
\end{figure*}

\citet{Vanderburg2014} present a method for mitigating or removing this noise, however it is not optimized for highly variable stars, e.g., the young stars of Upper Scorpius. The transit can still be identified in the \citet{Vanderburg2014} light curve, but systematic trends are present, including a discontinuity in the middle of the observations when \kepler\ changed the direction of its roll, and large changes in the point-to-point scatter over the observing window due to poor treatment of the thruster-fire systematics. Following \citet{Becker2015} and \citet{Mann2016} we extracted a new light curve by simultaneously fitting for low frequency variations from stellar activity, \kepler\ flat field, and the transits of \pname\ using a least-square minimization. Both stellar variability and the effect of errors in detector response were both modeled as splines as a function of time and centroid position with breakpoints every 0.2~days and 0.4\arcsec, respectively. Unlike \citet{2016ApJS..222...14V}, we did not apply separate systematics corrections to the first and second half of the \ktwo\ campaign, which removed the major discontinuity. The resulting light curve is relatively clear of visible systematic errors (see Figure~\ref{fig:lc}).

\subsection{Optical Spectrum from SNIFS}
We obtained an optical spectrum of \name\ on February 23, 2016 (UT) with the SuperNova Integral Field Spectrograph \citep[SNIFS,][]{Aldering2002,Lantz2004} on the University of Hawai'i 2.2m telescope on Maunakea. SNIFS covers 3200--9700\,\AA\ simultaneously with a resolution of $R\simeq$700 and R$\simeq$1000 in the blue (3200--5200\,\AA) and red (5100-9700\,\AA) channels, respectively. A single 710\,s exposure yielded S/N$=$80 per resolving element in the red channel. We also observed 5 spectrophotometric standards throughout the night for flux calibration. ThAr arcs were taken before or after each observation to improve the wavelength solution. Bias, flat and dark correction, as well as cosmic ray rejection, construction of the data cubes, and extraction of the one-dimensional spectrum are described in detail in \citet{Aldering2002}. The flux calibration is derived from the combination of the spectrophotometric standards and a model of the atmospheric absorption above Maunakea as described in \citet{Mann2015b}.

\subsection{NIR Spectrum from ARCoIRIS}
During the night of January 25, 2016 (UT), we acquired $z'YJHK$ spectra ($\simeq 0.8-2.45$\,\um) of \name\ using the ARCoIRIS spectrograph \citep{2014SPIE.9147E..2HS}, newly installed at the Cassegrain focus of the Blanco 4m telescope at the Cerro Tololo InterAmerican Observatory. ARCoIRIS is a fixed-format, cross-dispersed, long-slit spectrograph projected onto an HAWAII-2RG array having 18-$\mu$m pixels. We used its 110.5 l mm$^{-1}$ reflection grating and a 1.1$^{"}\times 28^{"}$ slit to obtain an approximate spectral resolution of $R\simeq3500$ across all six spectral orders. 

We placed the object at two widely separated positions along the slit, A and B, and took two series of ABBA nods, with per-nod position exposure times of 100\,s. Immediately afterwards, we took a similar series of ABBA nod observations for the A0V standard HD 146606. An accompanying Cu-He-Ar comparison lamp spectrum was also obtained for wavelength calibration. Data reduction of the target and calibrator was performed using the {\em SpeXtool} suite of IDL packages \citep[version 4.1,][]{Cushing2004} adapted for the data format and characteristics of the ARCoIRIS instrument ({\em priv comm.} Katelyn Allers\footnote{\href{https://www.dropbox.com/sh/wew08hcqluib8am/AAC6HiCeAVoiDPSe-MXRU2Cda?dl=0}{ARCoIRIS Spextool}}). Each difference (A-B) image was flat-fielded, wavelength calibrated, and extracted to produce a one-dimensional spectra. The telluric calibrator star (HD 146606) was used to telluric correct and flux calibrate the target spectrum \citep[employing the package {\em xtellcorr};][]{2004PASP..116..352V}. The final reduced and stacked spectrum has a peak S/N$>200$ per resolving element in the $H$ and $K$ bands.

\subsection{High Resolution NIR Spectrum}\label{sec:igrins}
We observed \name\ on the nights of January 30, February 26, March 28 and 29, 2016 (UT) with the Immersion Grating Infrared Spectrometer \citep[IGRINS,][]{Park2014} on the 2.7m Harlan J. Smith telescope at McDonald Observatory. IGRINS provides simultaneous $H$- and $K$-band (1.48-2.48\,\um) coverage with a resolving power of $R\simeq$45,000. Similar to the ARCoIRIS observations, the target was placed at two positions along the slit (A and B) and observed in an ABBA pattern. Each integration was 600\,s, which, when stacked, yielded a S/N$=$50-80 per resolving element near the center of each spectral order (at each of the four epochs).

The IGRINS spectra were reduced using version 2.1 of the publicly available IGRINS pipeline package\footnote{https://github.com/igrins/plp} \citep{IGRINS_plp}, which provides optimally extracted one-dimensional spectra of both the A0V standard and target. We used the A0V spectra to correct for telluric lines following the method outlined in \citet{Vacca2003}. Spectra without telluric correction were kept and used to improve the wavelength solution and provide a zero-point for the RVs. 

Radial velocities were determined from the IGRINS data as explained in \citet{Mann2016} and Mace et al. (in prep). In brief, we used the telluric lines like an iodine cell to lock the wavelength solution over epochs months apart, then cross-correlated the spectrum against 230 RV templates with spectral type M0-M6. The final assigned RV and error is the robust mean and standard error of the cross-correlation across all 230 templates. The absolute RV was taken to be the weighted mean of the four individual measurements, with an error limited by the zero-point error of $153$\,\mps. Relative RV errors are generally 40\,\mps, except for the first epoch, which had unusually high telluric contamination and lower S/N.

\subsection{Adaptive Optics Imaging and Aperture Masking}
On March 18 (UT), 2016 we obtained natural guide star adaptive optics (AO) imaging \citep{2000PASP..112..315W} and non-redundant aperture masking (NRM) of \name\ with the facility imager, NIRC2, on Keck II atop Maunakea. All observations were taken in vertical angle mode, using the smallest pixel scale ($9.952 \pm 0.002$ mas/pix). Imaging was taken with the $K'$ filter and masking with the 9-hole mask. After AO loops closed we took eight images, each with 20 coadds and an integration time of 0.5\,s per coadd. For NRM we took 10 interferograms, each with an integration of time of 20\,s and 1 coadd.

Each frame was linearized and corrected for geometric distortion using the NIRC2 solution from \citet{Yelda2010}. Images were dark-subtracted and flat-fielded. We interpolated over "dead" and "hot" pixels. Dead pixels were identified from superflats taken in 2006-2013 as any pixel with a response of $<$30\% in at least half of all superflats. Similarly, hot pixels were identified from a comparable set of superdarks as any pixel with $\ge$10 counts in at least half of the superdarks. Pixels with flux levels $>10\sigma$ above the median of the 8 adjacent pixels were flagged as cosmic rays or transient hot pixels and replaced with the median.

To detect faint and wide ($\gtrsim500$\,mas) companions in the AO images we subtracted an azimuthal median PSF model built from the smoothed PSF of \name. This added no additional noise at wide separations, but left the speckles in place, making it non-ideal for detecting close-in companions. To probe smaller inner working angles we instead constructed and subtracted the best-fitting PSF of another (single-star) target taken in the same night. We stacked all subtracted frames of \name, and identified companions by measuring the flux in 40\,mas (radius) apertures centered on every image pixel. We measured our detection limits from the standard deviation of the flux among all apertures in a 5-pixel annulus around the primary. We found no apertures that contained sufficient flux within the NIRC2 field of view to be considered an astrophysical source. 

The aperture masking observations use the complex triple product, or closure-phase, to remove non-common path errors introduced by atmospheric conditions and variable optical aberrations. To remove systematics, the observation of \name\ was paired with a calibration observation of USco J160933.8-190456, another member of Upper Scorpius \citep{2001AJ....121.1040P}. Binary system profiles can then be fit to the closure phases to produce separations and position angles and calculate contrast limits. The appendix of \citet{Kraus2008} contains a full explanation of the data reduction and binary profile-fitting for aperture masking data. No sources were detected in the masking data.

\begin{figure}
	\centering
	\includegraphics[width=0.46\textwidth]{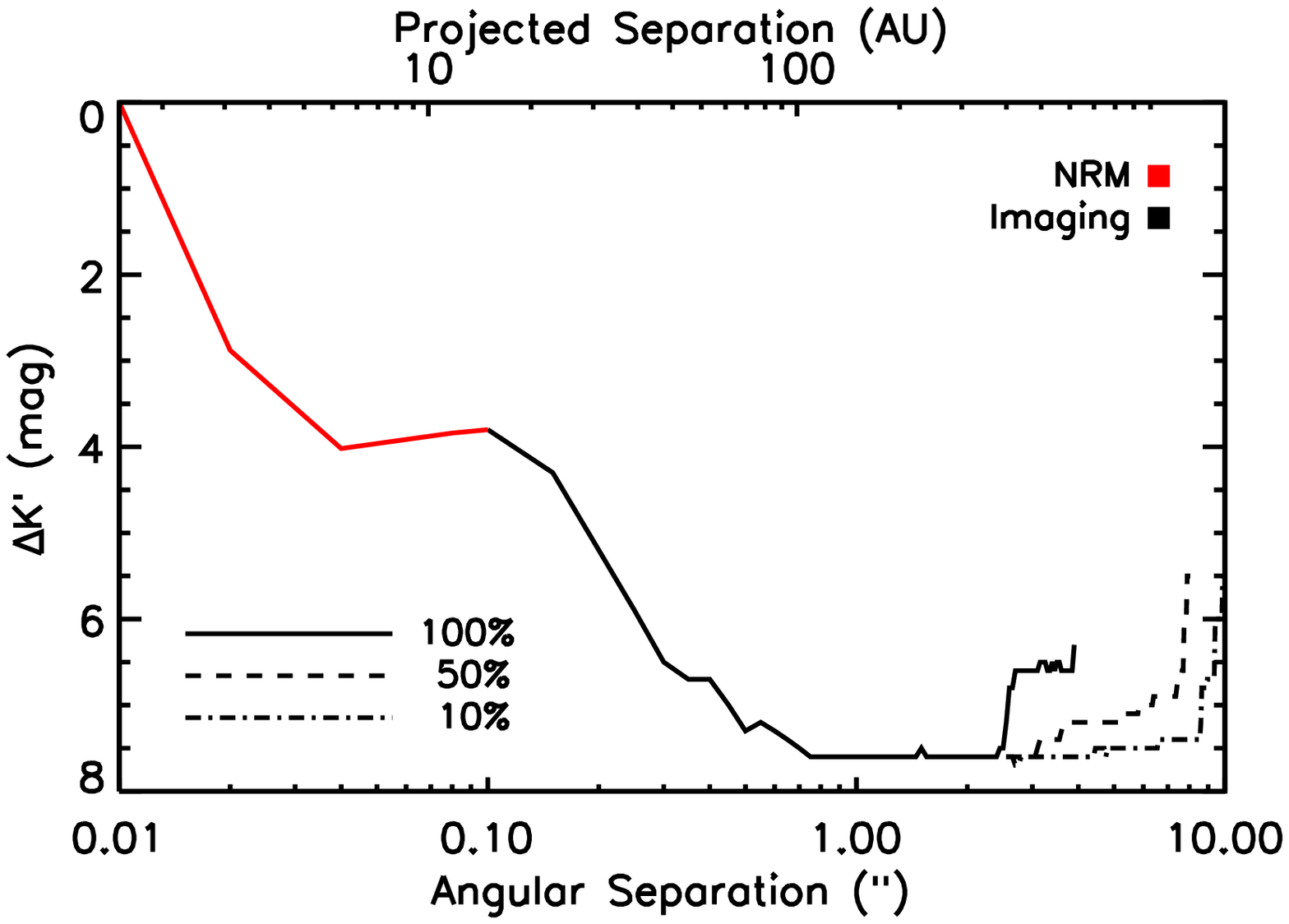} 
	\caption{Detection limits (5$\sigma$) for \name\ from our AO imaging and NRM interferometry as a function of separation. The top axis shows the separation in AU assuming a distance of 145\,pc. The region probed by non-redundant aperture masking is marked in red, while the region probed by imaging is in black. Owing to the finite chip size data is incomplete as a function of azimuthal angle at large separations. There we show the contrast curves with 100\% (solid line), 50\% (dashed), and 10\% (dotted-dashed) azimuthal completeness. }
	\label{fig:ao}
\end{figure}

The combination of the aperture masking and imaging observations excludes contributions of additional stars at separations 0.02-3" to the lightcurve. Figure~\ref{fig:ao} displays the imaging and masking contrast limits as a function of separation. Owing to the edges of the detector the azimuthal coverage is not complete outside of $\simeq$3\arcsec.

\subsection{Transit photometry from \mearth}
We observed two additional transits of \pname\ using the \mearth-North and \mearth-South arrays \citep{Nutzman:2008gf, 2012ApJ...747...35B, 2015csss...18..767I} on February 16, 2016 and on March 14, 2016 (UT). \mearth-North comprises eight 40-cm telescopes at Fred Lawrence Whipple Observatory on Mount Hopkins, Arizona. \mearth-South uses a nearly identical set of telescopes located at Cerro Tololo Inter-American Observatory (CTIO) in Chile. All telescopes use a $2048\times2048$ pixel CCD with pixels scales of 0.78\arcsec / pixel in the north and 0.84\arcsec / pixel in the south. The Schott RG715 filter was used for all observations \citep[see][for the filter profile and CCD transmission]{2016ApJ...818..153D}. 

All telescopes integrated for 60\,s for a cadence of $\simeq$90\,s per telescope. The first transit observation was only visible from \mearth-South, during which four telescopes simultaneously observed the second half of the transit, including $\gtrsim$1 hour after egress to fit for stellar variability. The second observation included the full transit, the first half of which was observed by four telescopes at \mearth-South and the second half by six telescopes in \mearth-North (including $\simeq$30\,m of simultaneous observations). Each array monitored \name\ for at least an hour of before or after the transit. In total more than 1200 photometric measurements were taken by \mearth\ during these transits.

\mearth\ also took low-cadence photometry of \name\ from January 26, 2016 to March 26, 2016 to constrain the long-term photometric variability. This long-term monitoring was done with one of the \mearth-South telescopes, which took two 60\,s exposures every 20-30\,minutes whenever the target was visible and the weather was amenable. 

MEarth data were reduced following the basic methodology from \citet{2007MNRAS.375.1449I} with additional steps detailed in the documentation of the fourth \mearth\ data release\footnote{\href{https://www.cfa.harvard.edu/MEarth/DR4/processing/index.html}{https://www.cfa.harvard.edu/MEarth/DR4/processing/index.html}}. This included corrections for second order extinction (color differences between target and comparison stars), meridian flips (when the target crosses the meridian the telescope rotates by 180 degrees relative to the sky, and reference stars fall on different parts of the detector), and stellar variability (fit from the transit and long-term monitoring). We also removed data points with anomalously high errors ($>0.8\%$, mostly taken during twilight). Lastly, we scaled the flux from the two telescope arrays to force agreement between the overlapping data. 

\section{Light Curve Analysis}\label{sec:lc}

\subsection{Transit Identification}\label{sec:ident}

The transit around \name\ was initially identified by \citet{2016ApJS..222...14V} as an Earth-sized planet orbiting an M dwarf (\teff=2890\,K, $R_*$=0.16$R_\odot$) every 5.425~days. \citet{2016ApJS..222...14V} assumed the host star was unreddened, causing the inferred \teff\ to be erroneously low. Further, the assumption that the star was on the main sequence led to an even more erroneously low inferred stellar radius. Thus the estimated radius of the planet was under-estimated as well.

The transit signal of \name\ was independently identified by the ZEIT project \citep{Mann2016} and the Mass-Radius Relation of Young Stars \citep{2015ApJ...807....3K}, both while searching for transiting/eclipsing systems in Upper Scorpius. The ZEIT search method is based on the box-least squares algorithm \citep{Kovacs2002}, but optimized for high amplitude rapid rotators. \name\ was the first Upper Scorpius planet candidate identified by our search, as it exhibits a comparatively high S/N ($\simeq25$) trapezoidal signal with a long-duration ($\simeq$4h) as expected for a planet around a still contracting, PMS star. 

\subsection{Transit fitting}\label{sec:transitfit}

We simultaneously fit the \ktwo\ and \mearth\ transit data with a Monte Carlo Markov Chain (MCMC) as described in \citet{Mann2016}, which we briefly summarize here. We used the \textit{emcee} Python module \citep{Foreman-Mackey2013} to fit the model lightcurves produced by the \textit{batman} package \citep{Kreidberg2015} using the \citet{MandelAgol2002} algorithm. Following \citet{Kipping:2010lr} we over-sampled and binned the model to match the 30\,minute \ktwo\ cadence. We used an unbinned model to fit the \mearth\ data due to the much lower integration time (60\,s). We sampled the planet-to-star radius ratio ($R_P/R_*$), impact parameter ($b$), orbital period ($P$), epoch of the first transit mid-point ($T_0$), bulk stellar density ($\rho_*$), and two limb darkening parameters ($q1$ and $q2$) for each of the two instruments (\mearth\ and \ktwo). At this young age it is likely that any eccentricity was dampened by the primordial disk \citep[e.g.,][]{2004ApJ...602..388T, 2007A&A...473..329C}, so we fix the eccentricity at zero. However, the eccentricity distribution of young planets has not been observationally constrainted; we discuss lifting this assumption in Section~\ref{sec:discussion}.

We assumed a quadratic limb darkening law and used the triangular sampling method of \citet{Kipping2013} in order to uniformly sample the physically allowed region of parameter space. We applied a prior on limb darkening derived from the \citet{2013A&A...553A...6H} atmospheric models, calculated using the LDTK toolkit \citep{2015MNRAS.453.3821P}, which enabled us to account for errors in stellar parameters and finite grid spacing. Stellar parameters input into LDTK are derived in Section~\ref{sec:params}. Errors on the limb darkening coefficients are broadened by a factor of two to account for model uncertainties (estimated by comparing limb darkening parameters from different model grids). The filter profiles and CCD transmission functions were taken from \citet{2016ApJ...818..153D} for the \mearth\ bandpass and from the \kepler\ science center\footnote{\href{http://keplergo.arc.nasa.gov/CalibrationResponse.shtml}{http://keplergo.arc.nasa.gov/CalibrationResponse.shtml}} for the \kepler\ bandpass. This yielded quadratic limb darkening coefficients of $\mu_1 = 0.4\pm0.1$ and $\mu_2 = 0.4\pm0.1$ for \kepler\ and $\mu_1 = 0.26\pm0.09$ and $\mu_2 = 0.4\pm0.1$ for \mearth. 

Our MCMC was allowed to explore $|b|<1+R_P/R_*$, $P$ from 0 to 70 days, $\rho_*$ from 0 to $\infty$, $R_P/R_*$ from 0 to 1, and $T_0$ from $\pm$3 days from the initial value, all with uniform priors. All parameters were initialized to the values from our BLS search (Section~\ref{sec:ident}), which are based on a Levenberg-Marquardt fit to the light curve \citep{Markwart2009}. MCMC chains were run using 150 walkers, each with 100,000 steps after a burn-in phase of 5,000 steps. 

\begin{figure}
	\centering
	\includegraphics[width=0.95\columnwidth]{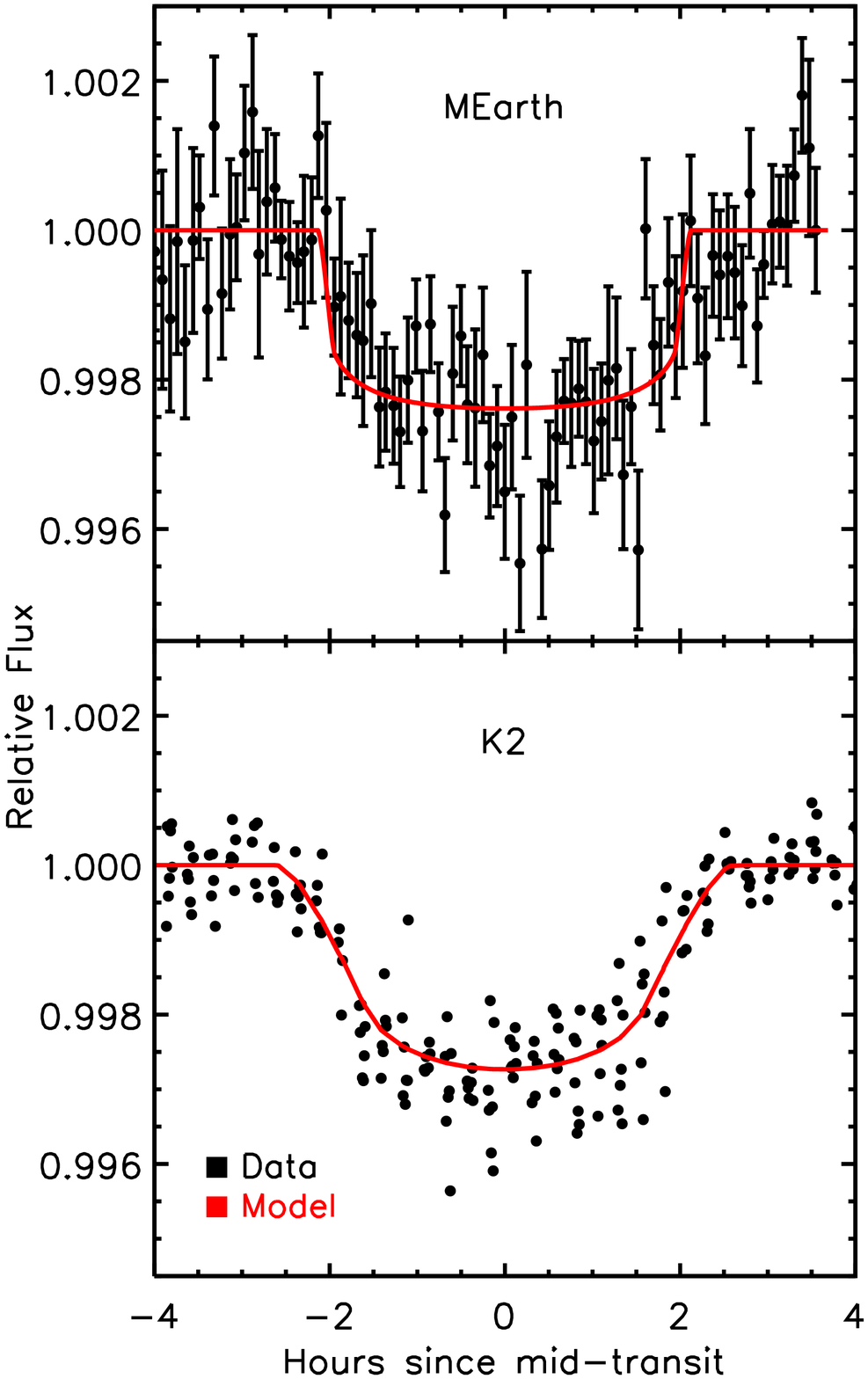} 
	\caption{Phase-folded light curve of \name's transit (black) from \mearth\ (top, binned) and \ktwo\ (bottom). The best-fit transit models are show in red. Owing to the large number of data points from \mearth\ we bin every 5\,min of data and show the median and $1\sigma$ scatter of points in each bin (unbinned data is used for MCMC fit). The \ktwo\ fit hasa longer ingress/egress because of the 30\,min integration time, which is accounted for in the model. Some systematics are present in the ingress of the \mearth\ transit, which we attribute to imperfect correction of stellar variability.}
	\label{fig:transitfit}
\end{figure}

\begin{figure}
	\centering
	\includegraphics[width=0.46\textwidth]{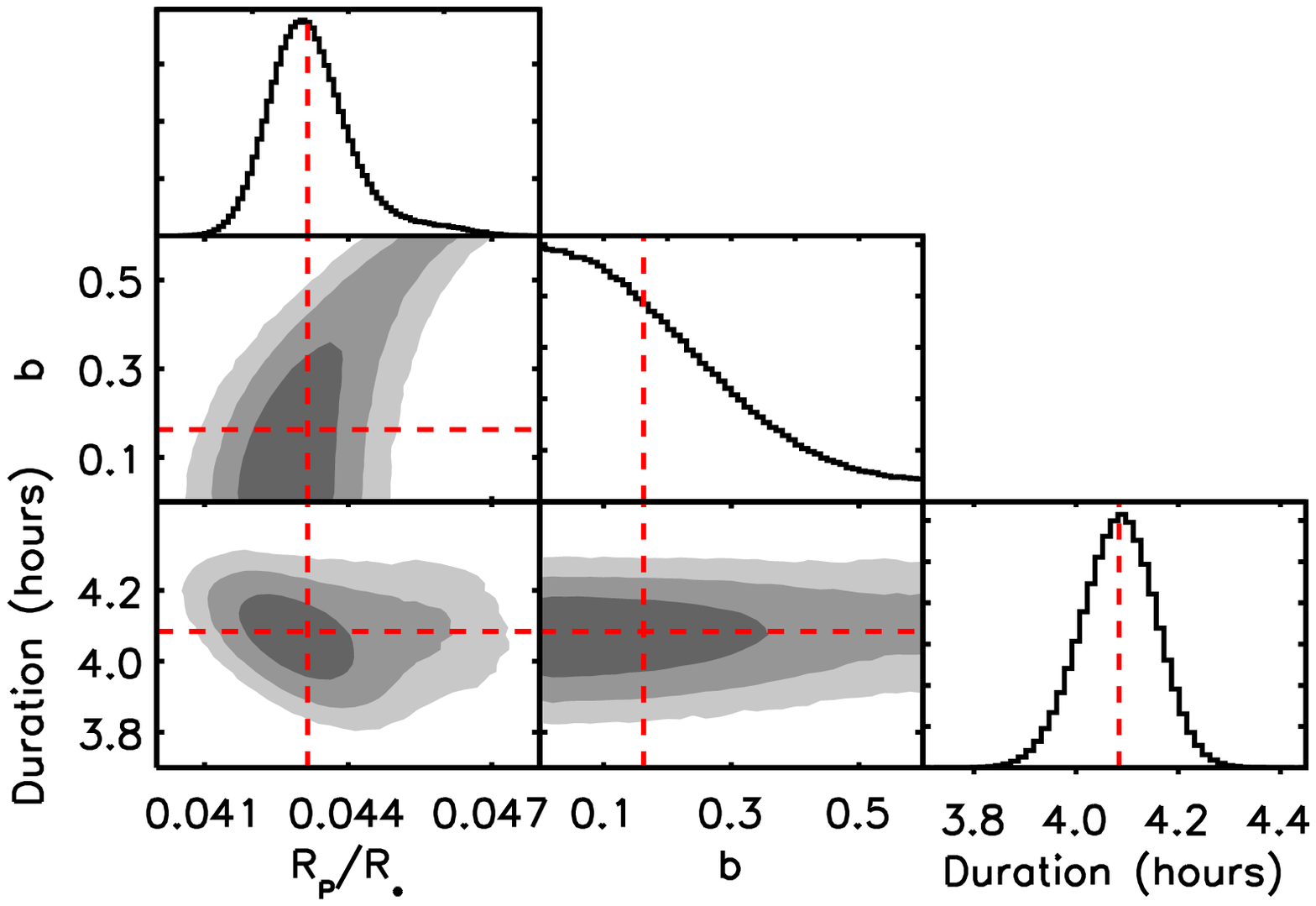} 
	\caption{Posteriors (histograms) and parameter correlations (contour plots) from our MCMC fit to the \mearth\ and \ktwo\ light curves. Median values for each parameter are marked with red dashed lines. Grey shading covers 67\%, 95\% and 99\%, from dark to light, of the MCMC posterior.}
	\label{fig:transitparams}
\end{figure}

We report the transit fit parameters in Table~\ref{tab:planet}. For each parameter we report the median value with the errors as the 84.1 and 15.9 percentile values (corresponding to 1$\sigma$ for Gaussian distributions). The model light curve with the best-fit parameters is shown in Figure~\ref{fig:transitfit} alongside the \ktwo\ and \mearth\ data. Some correlated errors are present in the \mearth\ light curve, primarily during ingress, which we attribute to imperfect correction of stellar variability and/or the planet crossing a spot. We also show posteriors and correlations for a subset of parameters in Figure~\ref{fig:transitparams}.

\floattable
\begin{deluxetable}{l l l l l l l l l l l }
\tablecaption{Transit Fit Parameters}
\tablewidth{0pt}
\tablehead{
\colhead{Parameter} & \colhead{Value} }  
\startdata
Period (days) & $5.424865^{+0.000035}_{-0.000031}$ \\
$R_P/R_*$ & $0.0432^{+0.0009}_{-0.0007}$ \\
T$_0$\tablenotemark{a} (BJD-2400000) & $56898.69288^{+0.00118}_{-0.00120}$ \\
Density ($\rho_{\odot}$) & $0.51^{+0.04}_{-0.07}$ \\
Impact Parameter & $0.16^{+0.19}_{-0.11}$ \\
Duration (hours) & $4.08^{+0.07}_{-0.07}$ \\
$a/R_*$ & $10.40^{+0.27}_{-0.50}$ \\
Inclination (degrees) & $89.1^{+0.6}_{-1.1}$ \\
Eccentricity & 0 (fixed) \\
$\omega$ (degrees) & 0 (fixed) \\
\hline
$R_P$\tablenotemark{b} (R$_\earth$)  & $5.04^{+0.34}_{-0.37}$ \\
\enddata
\tablenotetext{a}{BJD is in Barycentric Dynamical Time (TBD) format.}
\tablenotetext{b}{Planet radius is derived using our stellar radius (Section~\ref{sec:params}).}
\label{tab:planet}
\end{deluxetable}

The transit posterior favors a low ($<0.4$) impact parameter, although there is a tail in the distribution corresponding to higher impact parameter, lower $\rho_*$ ($<0.3$), and larger planet radius. This region of parameter space is not ruled out by our independent stellar parameters (see Section~\ref{sec:params}), so we did apply an additional constraint in the MCMC or remove these solutions from our transit-fit posterior. 

\section{Stellar Parameters}\label{sec:params} 

{\it Membership in Upper Scorpius:} The spatial position and kinematics of \name\ are consistent with co-motion of the star with the ensemble Upper Scorpius space velocity. We calculate a photometric distance to \name\ of 140$\pm$16\,pc using literature optical and NIR photometry, and the 10\,Myr isochrone from \citet{Chen2014}. This is consistent with the {\it Hipparcos} distances to high-mass members of Upper Scorpius \citep[145$\pm$15\,pc,][]{1999AJ....117..354D, 2011MNRAS.416.3108R}. Using this photometric distance, proper motions from UCAC4 \citep[-9.8$\pm$1.7, -24.2$\pm$1.8\,mas\,yr$^{-1}$][]{Zacharias:2012vn}, and the mean RV from our IGRINS observations, we calculate the Galactic space velocity of \name\ ($U$, $V$ ,$W$) = ($5.4\pm0.5$, $-15.8\pm2.2$, $-8.2\pm1.2$\,\kms). This is consistent with the kinematic models of \citet{Chen2014} and velocity dispersion of $\sim$2-3\,km/s \citep{Kraus2008b} for Upper Scorpius. Using the Bayesian method from \citet{2011MNRAS.416.3108R} and \citet{2015MNRAS.448.2737R} we calculate a probability of membership in Upper Scorpius of 96\% for \name.

\name\ also shows multiple indicators of youth. We measured the Na-8189 index, a well-calibrated gravity index \citep{slesnik06} to be 0.946$\pm$0.005, consistent with other late-type stars in Upper Scorpius and suggesting an age of 5-30\,Myr. \name\ shows a 24\,\um\ excess \citep{2012ApJ...758...31L} in Spitzer:MIPS observations, suggesting the presence of a disk, and hence an age for \name\ of $<$40\,Myr. Further, \name\ was already identified as a member of the Upper Scorpius subgroup by the presence of a strong Li 6708\,\AA\ line \citep[0.45$\pm$0.15\,\AA,][]{2001AJ....121.1040P}, an unambiguous indicator of youth for late-type stars.  

\begin{figure*}
	\centering
	\includegraphics[width=0.9\textwidth]{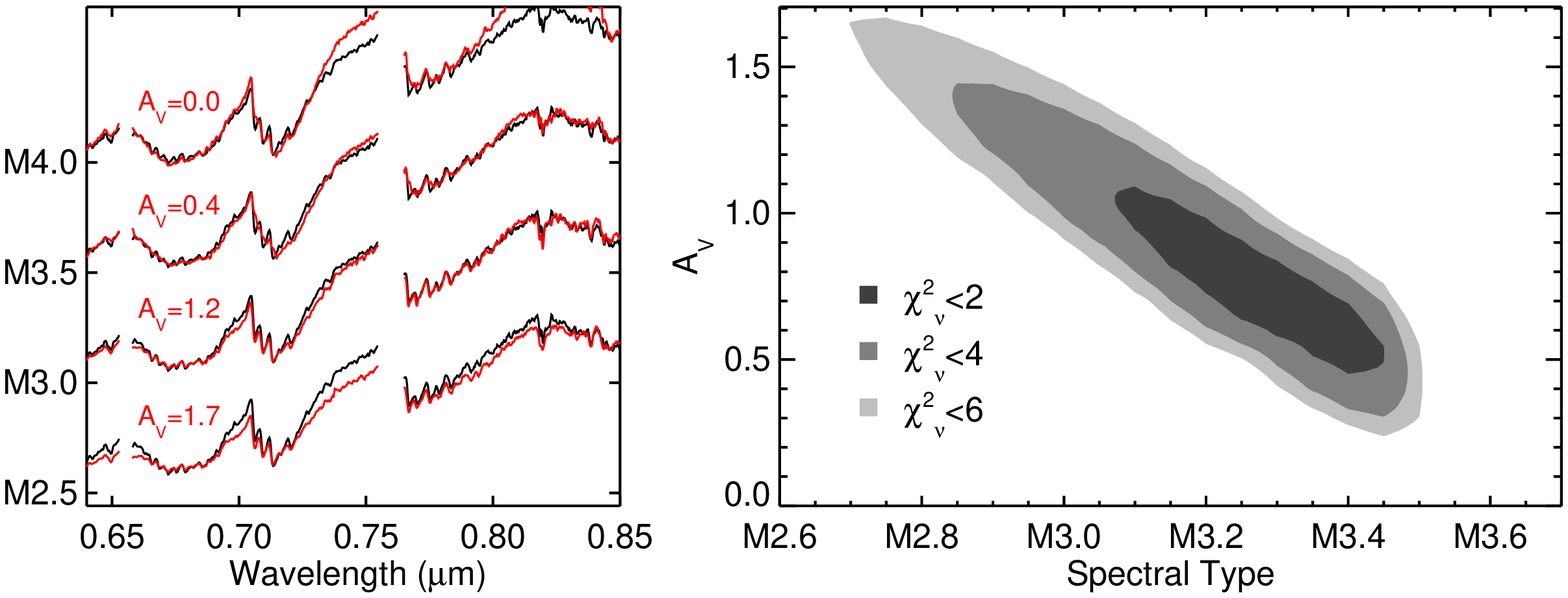} 
	\caption{Left: Optical spectrum of \name\ (black) compared to a M2.5-M4 young templates from \citet{2014ApJ...786...97H}, which are shown in red. For each template we found the best-fit \av\ value (lowest $\chi^2$). Right: the reduced $\chi^2$ (\rchisq) surface as a function of the template spectral type and reddening. This suggests a best-fit spectral type of M3.3$\pm$0.2 with an \av\ of 0.75$^{+0.21}_{-0.18}$.}
	\label{fig:reddening}
\end{figure*}

{\it Spectral type and reddening}: following \citet{2015ApJ...807....3K} and \citet{2016ApJ...816...69A} we simultaneously solved for spectral type and reddening (\av) to account for correlations between these parameters. We compared our optical spectrum of \name\ to a grid of 270 unreddened optical spectra of young stars from \citet{2014ApJ...786...97H}. For each template we computed the \av\ value that gives the best agreement between the spectrum of \name\ and that of the template using the reddening law from \citet{1989ApJ...345..245C} and masking out the H-$\alpha$ line and the strong O$_2$ tellurics. The resulting distribution of reduced $\chi^2$ (\rchisq) values yielded a spectral type of M3.3$\pm$0.2 with an \av\ of 0.75$^{+0.21}_{-0.18}$ (Figure~\ref{fig:reddening}). This spectral type error does not account for systematic errors in the spectral typing scheme (which can vary by 0.5-1 spectral subtype between methods), so we instead we adopted a more conservative M3.3$\pm$0.5. This does not affect the \av\ determination, but \av\ could be affected if there are systematic errors in the spectrophotometric calibration of our optical spectrum or the \citet{2014ApJ...786...97H} templates.

To test our sensitivity to our choice of templates, we repeat the above process with M dwarf spectral templates from \citet{Gaidos2014}. These were taken with the same instrument as our spectrum of \name, but the spectra are predominantly from old ($>1$\,Gyr) stars. The \citet{Gaidos2014} templates give a slightly earlier spectral type and higher reddening (M3.1, 0.83), but both are consistent with the values derived using templates from young stars. Comparison to the \citet{Gaidos2014} templates also give significantly higher \rchisq\ values than to those from \citet{2014ApJ...786...97H}, likely because of gravity-dependent differences in the spectrum.

{\it Effective Temperature}: We compared the unreddened spectrum to a grid of BT-SETTL CIFIST models\footnote{\href{https://phoenix.ens-lyon.fr/Grids/BT-Settl/CIFIST2011}{https://phoenix.ens-lyon.fr/Grids/BT-Settl/CIFIST2011}} \citep{Allard2011, Allard:2012fk}, masking our regions where models poorly reproduce observed spectra and accounting for small errors in the flux and wavelength solution as detailed in \citet{Mann2013} and \citet{Gaidos2014}. This method has been shown to reproduce \teff\ values for main-sequence M dwarfs derived from interferometry \citep{Boyajian2012}, but is poorly tested on PMS stars. However, we accurately reproduced the geometric \teff\ derived for the low-mass eclipsing binary USco CTIO5 \citep{2015ApJ...807....3K} suggesting our method yields reasonable \teff\ values even at young ages. To account for errors in reddening we repeated this process over the range of reddening values derived above. The effect of reddening is small, as the model comparison is driven primarily by the depth of the molecular bands instead of the overall spectral shape. We found a best-fit \teff\ of 3540$\pm$70~K.

{{\it Bolometric Flux}: we compiled optical $BVgri$ photometry from the ninth data release of the AAVSO All-Sky Photometric Survey \citep[APASS,][]{Henden:2012fk}, NIR $JHK_S$ photometry from The Two Micron All Sky Survey \citep[2MASS,][]{Skrutskie2006}, $griz$ photometry from the Sloan Digital Sky Survey \citep[SDSS,][]{Ahn:2012kx}, and $W1W2W3$ infrared photometry from the Wide-field Infrared Survey Explorer \citep[WISE,][]{Wright:2010fk}. We then scaled the (reddened) NIR and optical spectrum to the archival photometry following the procedure from \citet{Mann2015b}. The flux-calibrated spectrum is plotted in Figure~\ref{fig:spec}. We then unredden the calibrated spectrum. To calculate \fbol\ we integrated the spectrum over all wavelengths. As with \teff, this process was repeated over the range of \av\ values, which effectively tripled our error on \fbol. We found a best-fit for \fbol\ of $2.25 (\pm 0.26)\times10^{-10}$\,erg\,cm$^{-2}$\,s$^{-1}$.

\begin{figure}
	\centering
	\includegraphics[width=0.46\textwidth]{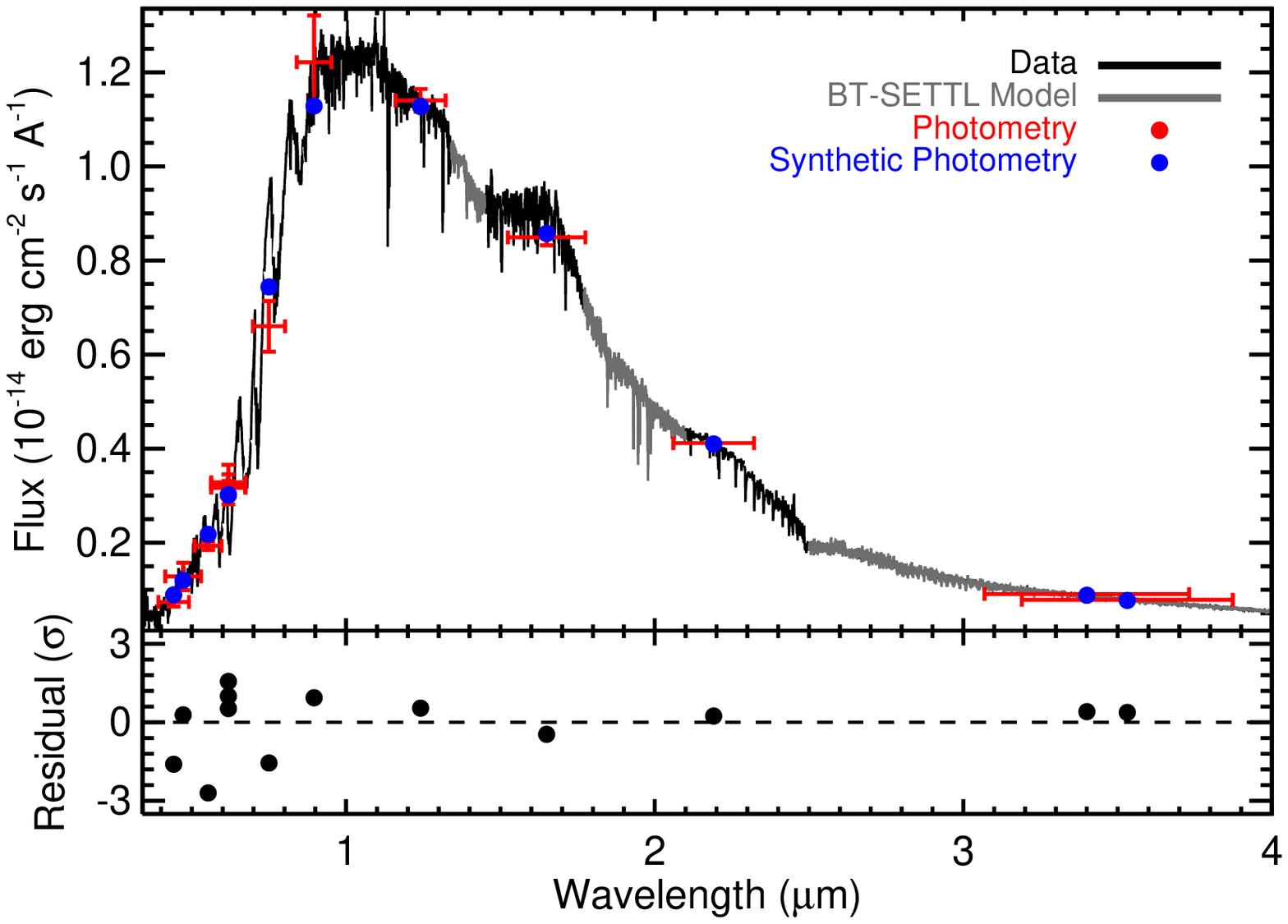} 
	\caption{Absolute flux calibrated spectrum of \name. Archival photometry is shown in red, with the horizontal error bars representing the effective width of the filter. Synthetic photometry derived from convolving the spectrum with the appropriate filter profile and zero-point \citep{2003AJ....126.1090C,Jarrett2011,Mann2015a} are shown in blue. We replace regions of high telluric absorption and those outside the range of our empirical spectra with an atmospheric model, which we show in grey. The spectrum and photometry shown here have not been corrected for reddening. The bottom pane shows the residual (photometry-synthetic) in units of standard deviations.}
	\label{fig:spec}
\end{figure}

{\it Stellar radius, mass and age}: we combined the distance to Upper Scorpius \citep[145$\pm$15\,pc,][]{1999AJ....117..354D} with our measured \fbol\ and \teff\ to calculate the stellar radius from the Stefan-Bolzman relation. This gives a radius of $1.02\pm0.13\,R_\odot$. We combined this with our transit-fit density from Section~\ref{sec:transitfit} to get a mass estimate of 0.55$^{+0.13}_{-0.14}\,M_\odot$. However, we can significantly improve on these parameters, and also estimate the age of the system using a grid of PMS stellar evolution models. 

We comparing our observables to two grids of PMS models computed with an updated version of the Dartmouth stellar evolution code \citep{Dotter2008, Feiden2012b}. A number of improvements to the code that allow more accurate computation of PMS stars are summarized in \citet{Feiden2016}. Germane to \name, one of the grids includes effects of magnetic inhibition of convection \citep{Feiden2012b, Feiden2013}, which relieves the observed age discrepancy between early- and late-type stars in Upper Scorpius \citep{2012ApJ...746..154P, 2015MNRAS.448.2737R}, yielding a consistent 9-10\,Myr median age for spectral types A through M \citep{Feiden2016}.

To infer the mass, radius, and age of \name, an MCMC method implemented with \textit{emcee} \citep{Foreman-Mackey2013} was used to sample the parameter space covered by our two model grids. The non-magnetic, standard model grid covered a mass range of 0.1--0.9 $M_\odot$ with a resolution of 0.02 $M_\odot$ and a metallicity range of $-0.5$ to $+0.5$~dex at a resolution of $0.1$~dex. The magnetic model grid covered a larger mass range of 0.1--1.7 $M_\odot$ with a resolution of 0.02 $M_\odot$, but the metallicity was restricted to [M/H]~$= 0.0$~dex. We explored mass, metallicity (for non-magnetic models), age, and distance to find an optimal fit to the observables: \teff\ and \fbol, using the likelihood function given in \citet{Mann2015b}. 

We applied a Gassian prior on distance \citep[145$\pm$15\,pc;][]{1999AJ....117..354D}, a prior on $\rho_*$ drawn from our transit fit posterior (Section~\ref{sec:transitfit}), and uniform priors on mass and age. For the non-magnetic models we applied a Gaussian prior on metallicity \citep[0.0$\pm$0.1~dex;][]{2011AJ....142..180B,2013prpl.conf1K086M}. To test the robustness of the results we also ran chains with a uniform prior on $\rho_*$ or distance (restricted to 10-1000\,pc) for each of the model grids.

\floattable
\begin{deluxetable}{L | L L | L L L L L L L}
\tabletypesize{\scriptsize}
\tablecaption{Stellar Fit Parameters \label{tab:stellar}}
\tablewidth{0pt}
\tablehead{
\colhead{Models} & \multicolumn{2}{c}{Priors} & \multicolumn{6}{c}{Parameters} \\
\colhead{} & \colhead{Distance (pc)} & \colhead{$\rho_*$ ($\rho_\odot$)} & \colhead{$R_*$~($R_\odot$) } & \colhead{$M_*$~($M_\odot$) } & \colhead{$L_*$~($L_\odot$) }& \colhead{Age (Myr)} & \colhead{[Fe/H]} & \colhead{Distance (pc)} \\
} 
\startdata
\rm{No\ model}\tablenotemark{a} & \nodata & \nodata & $  1.02^{+0.13}_{-0.13}$ & $  0.55^{+0.13}_{-0.15}$ & $  0.15^{+0.03}_{-0.03}$ & \nodata & \nodata & \nodata \\
\hline
 & 145$\pm$15 & \rm{uniform} &   1.01^{+0.18}_{-0.17}$ &$  0.56^{+0.09}_{-0.09}$ &$  0.14^{+0.05}_{-0.04}$ &$ 10.80^{+8.93}_{-4.71}$ &0~\rm{(fixed)} & $  143.2^{+ 21.0}_{- 21.6}$\\
\rm{Magnetic} & \rm{uniform} & $0.51^{+0.04}_{-0.07}$\tablenotemark{b} &   1.06^{+0.09}_{-0.08}$ &$  0.58^{+0.10}_{-0.10}$ &$  0.16^{+0.05}_{-0.04}$ &$  9.31^{+1.11}_{-1.48}$ &0~\rm{(fixed)} & $  152.8^{+ 25.2}_{- 23.5}$\\
 & 145$\pm$15 & $0.51^{+0.04}_{-0.07}$\tablenotemark{b} &   1.05^{+0.07}_{-0.07}$ &$  0.56^{+0.09}_{-0.09}$ &$  0.15^{+0.03}_{-0.03}$ &$  9.34^{+1.07}_{-1.26}$ &0~\rm{(fixed)} & $  148.0^{+ 15.6}_{- 15.9}$\\
\hline
 & 145$\pm$15 & \rm{uniform} &   0.97^{+0.18}_{-0.16}$ &$  0.42^{+0.10}_{-0.08}$ &$  0.14^{+0.05}_{-0.04}$ &$  6.06^{+6.18}_{-2.83}$ &$  0.01^{+0.13}_{-0.14}$ &$  140.1^{+ 20.1}_{- 20.7}$\\
\rm{Non-magnetic} & \rm{uniform} & $0.51^{+0.04}_{-0.07}$\tablenotemark{b} &   0.94^{+0.08}_{-0.12}$ &$  0.42^{+0.10}_{-0.09}$ &$  0.12^{+0.04}_{-0.04}$ &$  6.53^{+1.56}_{-1.00}$ &$  0.00^{+0.13}_{-0.14}$ &$  135.7^{+ 25.2}_{- 24.3}$\\
 & 145$\pm$15 & $0.51^{+0.04}_{-0.07}$\tablenotemark{b} &   0.95^{+0.07}_{-0.09}$ &$  0.43^{+0.09}_{-0.07}$ &$  0.13^{+0.03}_{-0.03}$ &$  6.57^{+1.44}_{-0.96}$ &$  0.01^{+0.13}_{-0.14}$ &$  139.5^{+ 18.0}_{- 19.2}$\\
\hline
 \enddata
\tablenotetext{a}{Radius from Stefan-Bolzman relation, mass from radius and transit-fit density.}
\tablenotetext{b}{Density prior taken from transit-fit posterior (Section~\ref{sec:transitfit}).}
\end{deluxetable}

The MCMC simulation was set up with 300 walkers at random initial starting positions and was allowed to run over 1000 iterations following a burn-in phase of 250 iterations. Convergence was diagnosed through a combination of visually monitoring trace plots for all 300 chains, monitoring the median acceptance fraction among all chains (between 25\% and 50\%), and by monitoring the auto-correlation time for individual chains.  A summary of all MCMC results is given in Table~\ref{tab:stellar}. Quoted values represent the median of the posterior and quoted uncertainties are the 68\% Bayesian credible intervals. 

All chains produced consistent stellar radii, luminosities, and masses. This is in large part a consequence of the observables and constraints; \teff, \fbol, and distance uniquely determine the stellar radius, luminosity, and, when combined with the transit-fit density, the mass. Thus when the stronger distance and density priors were used the resulting radius/luminosity/mass is relatively independent of the model grid used. It is encouraging that even when using a uniform distance prior yields a distance consistent with the previously established vale for Upper Scorpius. Further, all fits to the magnetic models give an age consistent with the value estimated for high-mass members of Upper Scorpius \citep{2012ApJ...746..154P}. 

Using a uniform prior on $\rho_*$ results in a negligible change in mass and radius, although with much larger errors. This suggests the accuracy of our stellar parameters (but not the precision) is insensitive to the assumption of zero orbital eccentricity for \pname. We show a comparison of the model-based and transit-fit densities in Figure~\ref{fig:density}. Because the constraints on $\rho_*$ are relatively weak from the model comparison alone, we cannot make definitive statements about the orbital eccentricity of the system from the data alone, and instead rely on the physical argument that a planet migrating via interaction with the disk should have $\simeq$0 orbital eccentricity. 

\begin{figure}
	\centering
	\includegraphics[width=0.46\textwidth]{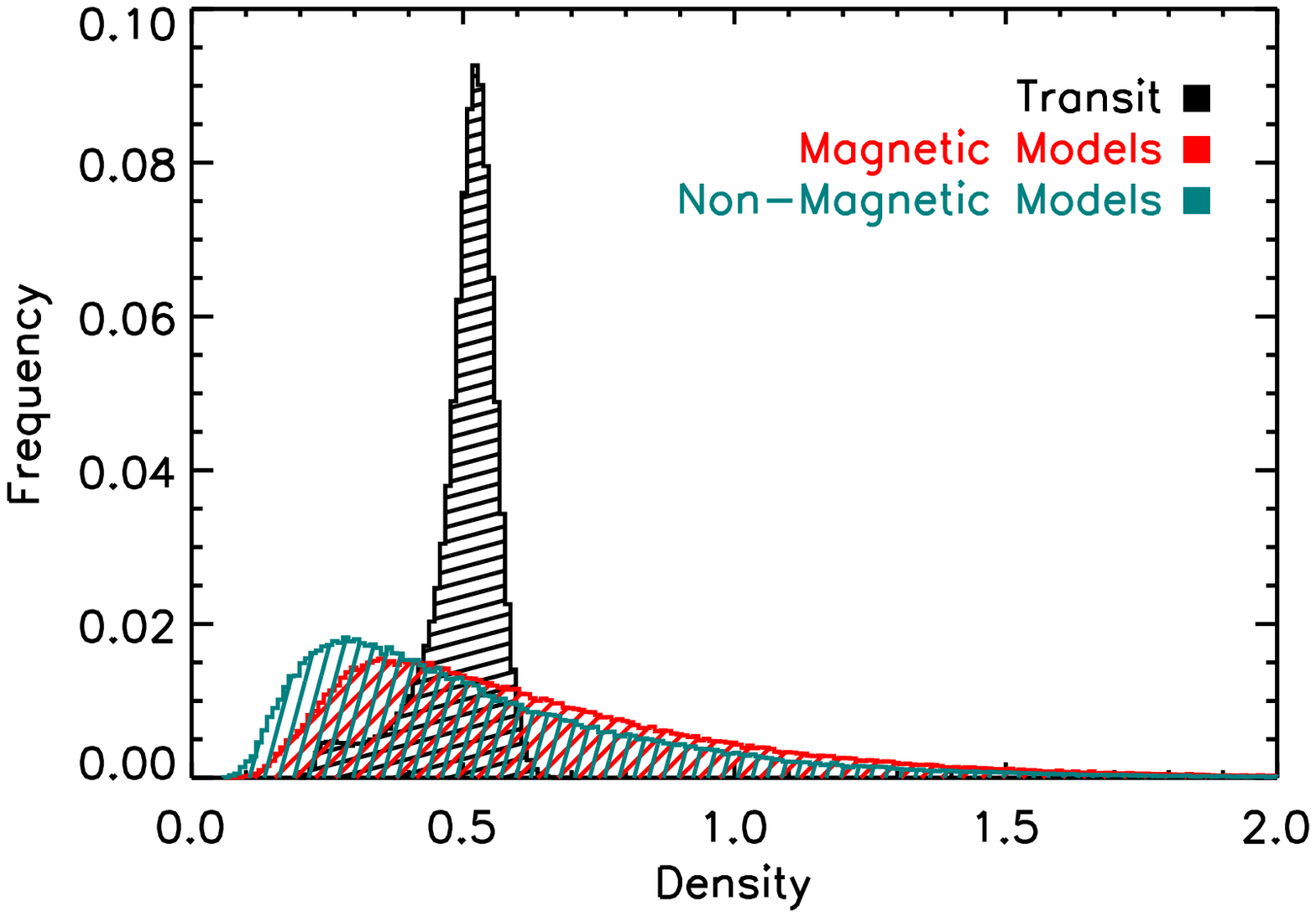} 
	\caption{Stellar density ($\rho_*$) from our transit fit (black, Section~\ref{sec:transitfit}) assuming $e=0$ compared to that from our MCMC model comparison. Red indicates the result when using magnetic models, while teal shows the result from models without correcting for magnetic fields. Both model results shown here use uniform priors on stellar density. }
	\label{fig:density}
\end{figure}

The non-magnetic models favor a lower mass and radius, and younger age. Magnetic models more accurately reproduce the known distance and age of Upper Scorpius. We therefore adopted parameters from the chain utilizing magnetic models with the distance prior from \citet{1999AJ....117..354D} and the density prior from our transit fit, which we use for the rest of the analysis. 

{\it Rotation period}: we computed the Lomb-Scargle periodogram of the \ktwo\ light curve prior to removing the stellar variability. A strong signal is apparent at 6.29~days, which we attribute to spot coverage and the rotation period. We estimate an error of 0.17~days on the rotation period from the width of the peak in the periodogram power spectrum. The same period was found from the \mearth\ long-term monitoring (6.27~days).

{\it Rotation velocity}: We determined \vsini\ using our high-resolution IGRINS spectrum. We first determined the instrumental profile/resolution by fitting the telluric spectrum derived from the A0V standard (see Section~\ref{sec:obs}) with a series of Gaussian profiles and assuming telluric lines have negligible intrinsic width compared to the instrument resolution. Orders with $<4$ strong ($>3\%$ depth) telluric lines were ignored. We assumed the resolution varies linearly within an order and smoothly in between orders. Our derived instrumental broadening was 0.3-0.5\,\AA\ (full-width half-max), consistent with the previously measured resolution of the spectrograph. 

We then compared our IGRINS spectrum of \name\ to the best-fit BT-SETTL model derived from the moderate resolution spectra above. We broadened the model first using the instrumental profile derived from our telluric fit, then as a function of \vsini\ using the IDL code \textit{lsf\_rotate} \citep{Gray1992, 2011ascl.soft09022H}. We included seven nuisance parameters to handle normalization of the observed spectrum, small errors in wavelength calibration, and missing or overly deep/shallow lines in the atmospheric model. Orders with S/N$<20$ were removed. Each order was fit separately, each time adjusting \vsini\ and the nuisance parameters to minimize the difference between the model and IGRINS spectrum. We adopted the mean and standard error of the \vsini\ determinations across all orders as the final value and error: $8.2\pm1.8$\,\kms. 

{\it Sky projected stellar inclination}: The combination of \vsini, rotation period, and stellar radius enabled a calculation of the (sky-projected) rotational inclination ($i_*$) of \name. We first calculated the equatorial velocity 
$V_{\rm{eq}} = \frac{2\pi R_*}{P_{\rm{rot}}}$, where $P_{\rm{rot}}$ is the stellar rotation period measured above, which yielded a velocity of $8.6\pm0.7$\,\kms. We assumed effects of differential rotation are encapsulated in our $P_{\rm{rot}}$ measurement error, although this depends on where on the star the spots are located. We converted \vsini\ and $V_{\rm{eq}}$ to a posterior in $cos(i_*)$, which handles regions of the posterior where \vsini$>V_{\rm{eq}}$ \citep[see][ for more details]{Morton2014b}. The resulting posterior gives a lower limit on stellar inclination of $i_*>63^{\circ}$ at 68.3\% (1$\sigma$), suggesting the planetary orbit is not highly misaligned with the stellar rotation. 

A summary of all derived stellar parameters and errors is given in Table~\ref{tab:params}.

\floattable
\begin{deluxetable}{l c l }
\tabletypesize{\scriptsize}
\tablecaption{Parameters of \name\ \label{tab:params}}
\tablewidth{0pt}
\tablehead{
\colhead{Parameter} & \colhead{Value} & \colhead{Source}
}
\startdata
\multicolumn{3}{c}{\hspace{1cm}Identifiers} \\
K2-33 && {\it K2} \\
EPIC~205117205 && EPIC\\
USco J161014.7-191909 && \citet{2002AJ....124..404P}\\
2MASS J16101473-1919095 && 2MASS \\
\hline
\multicolumn{3}{c}{\hspace{1cm}Astrometry} \\
$\alpha$ R.A. (hh:mm:ss J2000) & 16:10:14.73 & EPIC\\
$\delta$ Dec. (dd:mm:ss J2000) & -19:19:09.38 & EPIC\\
$\mu_{\alpha}$ (mas~yr$^{-1}$) & $-$9.8 $\pm$ 1.7 & UCAC4 \\
$\mu{\delta}$ (mas~yr$^{-1}$) & $-$24.2 $\pm$ 1.8 & UCAC4 \\
\hline
\multicolumn {3}{c}{\hspace{1cm}Photometry} \\
$B$ (mag) & 17.353 $\pm$ 0.138 & APASS \\
$g'$ (mag) & 16.386 $\pm$ 0.076 & APASS \\
$r'$ (mag) & 14.860 $\pm$ 0.072 & APASS \\
$i'$ (mag) & 13.360 $\pm$ 0.125 & APASS \\
$z'$ (mag) & 12.613 $\pm$ 0.004 & SDSS \\
$J$ (mag) & 11.095 $\pm$ 0.023 & 2MASS\\
$H$ (mag) & 10.332 $\pm$ 0.021 & 2MASS\\
$K_s$ (mag) & 10.026 $\pm$ 0.019 & 2MASS\\
$W1$ (mag) & 9.890 $\pm$ 0.023 & ALLWISE \\
$W2$ (mag) & 9.762 $\pm$ 0.021 & ALLWISE\\
$W3$ (mag) & 9.610 $\pm$ 0.049 & ALLWISE \\
\hline
\multicolumn{3}{c}{\hspace{1cm} Kinematics and Distance} \\
Barycentric RV (\kms) & -6.70 $\pm$ 0.15 & This paper \\
$U$ (\kms) & $ -5.4\pm  0.5$ & This paper \\
$V$ (\kms) & $-15.8\pm  2.2$ & This paper \\
$W$ (\kms) & $ -8.2\pm  1.2$ & This paper \\
Distance (pc) &   $145 \pm 15$ & \citet{1999AJ....117..354D}\\
\hline
\multicolumn{3}{c}{\hspace{1cm} Physical Properties} \\
\av & 0.75$^{+0.21}_{-0.18}$ & This paper \\
Spectral Type &  M3.3 $\pm$ 0.5 & This paper\\
Rotation Period (days) &  6.29 $\pm$ 0.17 & This paper \\
\teff\ (K) & 3540 $\pm$ 70 & This paper\\
\fbol\ ($10^{-10}$\,erg\,cm$^{-2}$\,s$^{-1}$) & $2.25 \pm 0.26$ & This paper \\
$M_*$ ($M_\odot$) & 0.56$^{+0.09}_{-0.09}$  & This paper\tablenotemark{a} \\
$R_*$ ($R_\odot$) & 1.05$^{+0.07}_{-0.07}$& This paper\tablenotemark{a} \\
$L_*$ ($L_\odot$) & 0.15$^{+0.03}_{-0.03}$ & This paper\tablenotemark{a} \\
Age (Myr) & 9.3$^{+1.1}_{-1.3}$ & This paper\tablenotemark{a}\\
\vsini\ (km~s$^{-1}$) & $8.2\pm1.8$  & This paper \\
$i_*$ (degrees) & $>63$ & This paper \\
\enddata
\tablenotetext{a}{Adopted parameters from our model comparison; see Table~\ref{tab:stellar} for more information.}
\end{deluxetable}

\section{False Positive Analysis}\label{sec:fpp}

\subsection{Background Eclipsing Binary}\label{sec.background}

We calculated a posterior probability that an unrelated, unresolved background source (i.e., an eclipsing binary) is responsible for the transit signal. The procedure is described in \citet{Gaidos2016}, and only summarized here. The Bayesian probability was calculated with a prior based on a model of the background stellar population drawn from TRILEGAL version 1.6 \citep{Vanhollebeke2009}. The likelihood is calculated from the observational constraints, i.e.,: (1) a background star must be bright enough to produce the transit signal given a maximum 50\% eclipse depth; (2) the density of the star must be consistent with the measured transit duration; and (3) the star must not be visible in our NIRC2 AO imaging and NRM interferometry (Section~\ref{sec:obs}).

Stars were selected from a TRILEGAL-generated synthetic catalog of 25,348 stars to $K_p = 22$ in a field of 10 sq. deg at the coordinates of \name. All standard settings were used except the extinction at $\infty$ was set to $A_V = 0.894$ based on the map of \citet{1998ApJ...500..525S}. We randomly placed stars at locations in a circular field within 4 \kepler\ pixels (16\arcsec) around \name. Stars were discarded if their position and magnitude are ruled out by the detection limits from our imaging/NRM, or they are too faint to produce the observed signal. We weighted each remaining star by the probability that an eclipsing binary with a period of 5.425~days would yield a duration consistent with our transit fit ($\simeq$4.1\,hours). 

The final false-positive probability is sensitive to the assumed binary eccentricity distribution. It is $< 3 \times 10^{-7}$ for all reasonable choices, but is essentially zero if the orbits are near-circular, as is expected for short-period binaries; the long transit duration could only be produced by a giant star, but these are bright and ruled out by the lack of detections in our AO data. Background stars fainter than \name\ will be denser dwarf stars. 

\subsection{Companion Eclipsing Binary}

We next considered the possibility that the transit signal is due a physically associated companion system, i.e., a very-low mass eclipsing binary (EB). Since the maximum transit depth of an EB is 50\%, the contrast ratio of such a companion must be $\Delta K_p > 5.7$, and thus absolute $M_{K_p} > 14.0$. Such a system cannot be excluded if it is within 0.2" of the primary, or a projected separation of $< 28$~AU. According to an 11-Myr isochrone  generated by the Dartmouth Stellar Evolution program, such a system would have to be substellar, i.e., a pair of eclipsing brown dwarfs or self-luminous young giant planets. Moreover splitting the light curve into even and odd transits shows that the ``primary'' and ``secondary'' transits have equal depths ($0.27 \pm 0.05$ and $0.25 \pm 0.04$\%), so such a system must consist of equal-mass components. This is inconsistent with the ``flat-bottomed'' shape of the transit light curve (Figure~\ref{fig:transitfit}).

\subsection{Eclipsing Binary}

To confirm that the transiting body is non-stellar, we use the RVs from our IGRINS spectra to put an upper limit on the mass of \pname. We fit the RVs assuming a circular orbit and locking the period and argument of periapsis from the transit fit (Section~\ref{sec:transitfit}) and assuming the mass derived for the host star from our model interpolation (Section~\ref{sec:params}). The RVs rule out companion masses above 3.7 Jupiter masses ($M_J$) at $3\sigma$ (Figure~\ref{fig:rvs}). The constraints are tighter ($<2.8M_J$) if we remove the IGRINS epoch with high telluric contamination (see Section~\ref{sec:obs}). If we loosen our assumptions about the orbital eccentricity then the maximum mass increases to 5.4$M_J$. In all cases the RVs exclude any brown dwarf or stellar companion ($>13M_J$) with an orbital period matching the transit signal.

\begin{figure}
	\centering
	\includegraphics[width=0.46\textwidth]{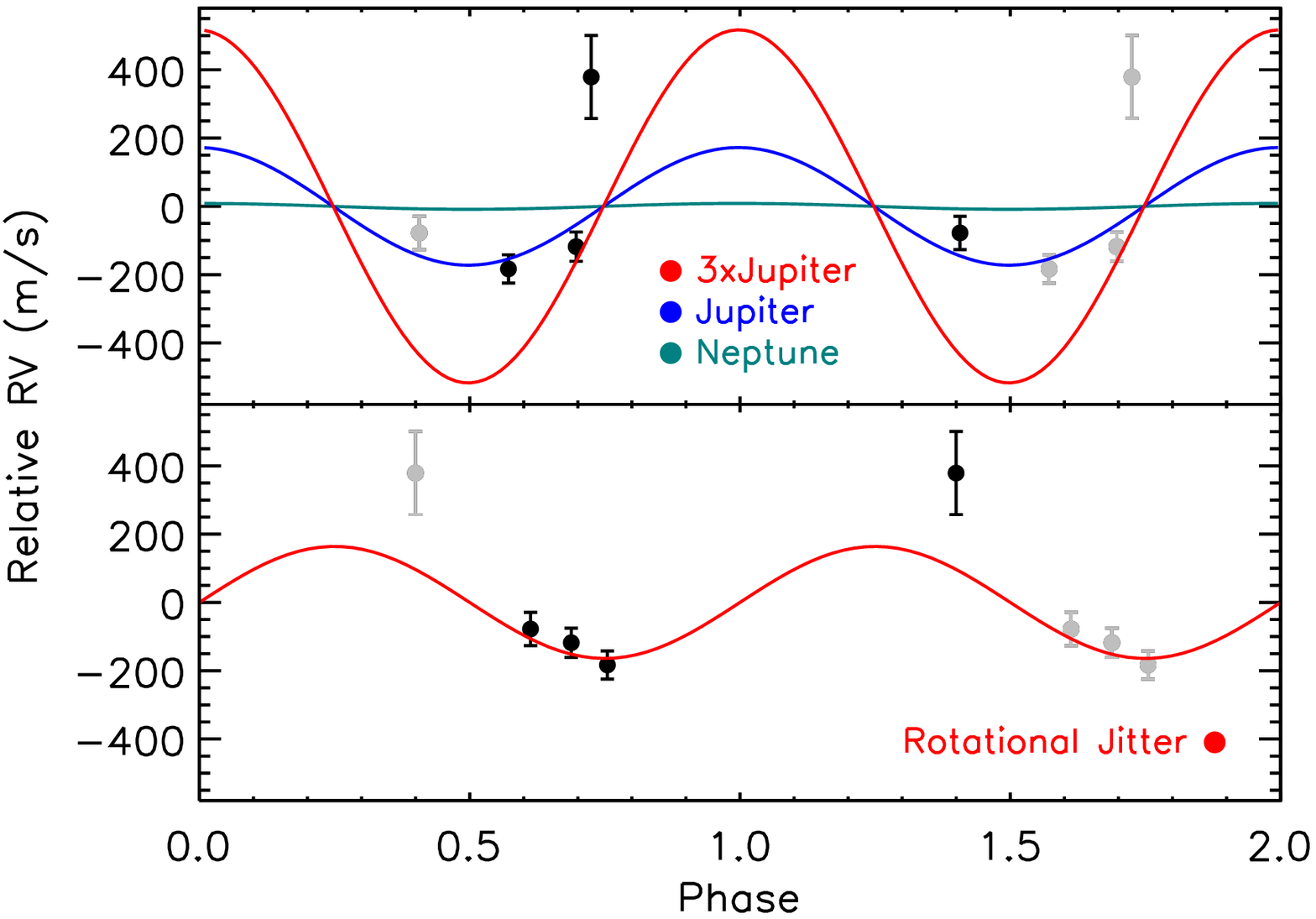} 
	\caption{Radial velocities derived from our IGRINS spectra, phased to the transit-based orbital period (5.42~days, top) and stellar rotation period (6.29\, days, bottom). Duplicate measurements are shown in grey. The expected RV amplitudes (assuming circular orbits) for Neptune-mass, Jupiter-mass, and $3 M_J$-mass planets at this orbital period are shown as teal, blue, and red lines on the top panel. An estimate of the spot-induced RV jitter, derived from the \vsini\ and variability in \ktwo\ data, is shown in the bottom panel in red.}
	\label{fig:rvs}
\end{figure}

\subsection{Stellar Variability}

Spots and plages on the photosphere combined with stellar rotation create 1-3\% variations in the light curve of \name. The amplitude of this variation is roughly an order of magnitude larger than the transit depth ($\simeq$0.26\%, Figure~\ref{fig:transitfit}). Fortunately, spots create a characteristic shape in the light curve curve (smoothly varying) and duration ($\simeq$ half the rotation period) that differs from a transit (trapezoidal shape and a duration of hours). This makes them easy to differentiate in most stars. However, improper removal of the more complicated spot patterns on young stars can sometimes generate transit-like signals over short (days or weeks) timescales. Our BLS search identified many such systems; one of which we show in Figure~\ref{fig:badlc}

\begin{figure}
	\centering
	\includegraphics[width=0.46\textwidth]{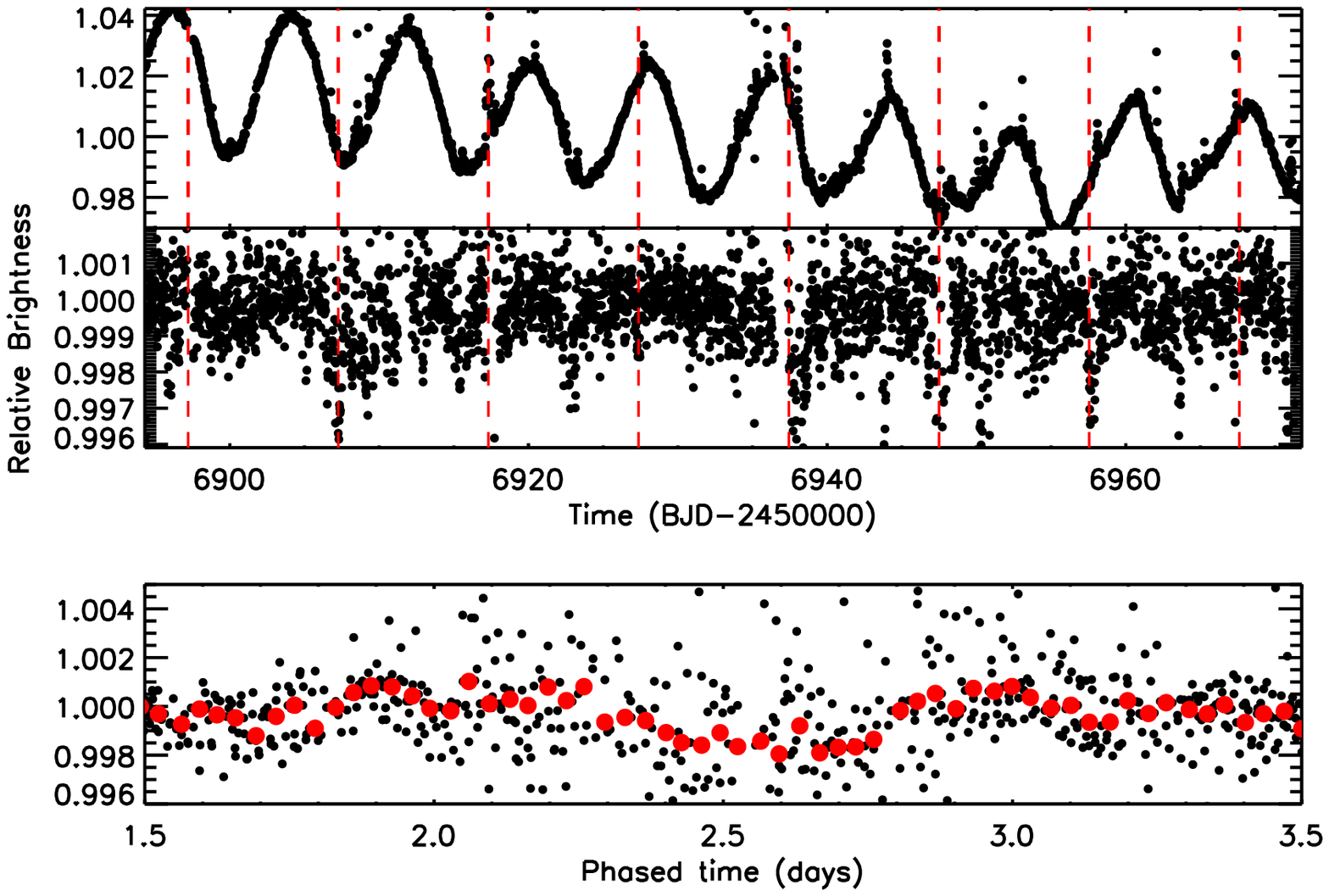} 
	\caption{Light curve of a star in Upper Scorpius taken by the \ktwo\ mission, with the top two panels following the layout of Figure~\ref{fig:lc}. The bottom panel shows the light curve folded to the highest power period identified from the BLS, with data binned every 20\,min in red. Poor removal of the flares, stellar variability, and \ktwo\ drift creates systematic noise in the flattened light curve. When folded this can look like a weak transit. However, the individual transits have inconsistent depths and durations and the out-of-transit light curve contains numerous residual variations. This was identified as a candidate by our BLS search but subsequently identified as a false positive.}
	\label{fig:badlc}
\end{figure}

The combination of our \mearth\ and \ktwo\ light curves demonstrate that the transit signal cannot be caused by stellar variability. While spot patterns can be stable on multi-year baseline in old M dwarfs \citep{Newton2016}, spot evolution can been seen even over the 80\,day observing window for Upper Scorpius members (e.g., Figure~\ref{fig:lc}, \ref{fig:badlc}). If the transit signal was an artifact of stellar activity signals the transit depth and shape would change or disappear over the \ktwo\ observing window and between the \ktwo\ and \mearth\ observations. To test this we fit each transit individually as in Section~\ref{sec:transitfit}, but locking the period to the previously derived value. We found that all transits yield consistent depths and transit durations, including \mearth\ data taken 1.5 years after the \ktwo\ observations. Further, stellar signals are generally wavelength dependent, and the separate fits to the \ktwo\ ($\lambda_{\rm{mean}}\simeq6400$\,\AA) and \mearth\ ($\lambda_{\rm{mean}}\simeq8250$\,\AA) light curve give consistent parameters. 

\subsection{Debris Disk?}

Many stars in Upper Scorpius exhibit excess emission at infrared and millimeter wavelengths indicative of cooler, circumstellar material, i.e., dusty primordial or debris disks in different stages of their evolution \citep{2012ApJ...758...31L}. \name, may have excess emission at 24\,\um\ (measured by the MIPS instrument on {\it Spitzer}), but shows no significant excess in any of the WISE bands or the {\it Spitzer} 8 and 16\,\um\ bands \citep{2009ApJS..181..197C, 2012ApJ...758...31L}. We confirmed this by comparing the {\it Spitzer} measurements from \citet{2012ApJ...758...31L} to our estimate of the photospheric flux (see Section~\ref{sec:params}), extrapolated to 30\,\um\ using a PHOENIX BT-SETTL model (Figure~\ref{fig:IR}). This excess is consistent with the WISE upper limit at 22\,\um\ (W4 channel). This excess corresponds to that of debris disks, not an evolved disk, according to the classification of \citet{2012ApJ...758...31L}.

The lack of a detectable excess at wavelengths $<12$\,\um\ suggests that any disk, if it exists, must lack significant material warmer than 300\,K, i.e., a central hole, or a drop in the emissivity of the grains close to the star. Using the value for $L_*$ estimated in Section~\ref{sec:params}, we estimated that the hole extends to $>0.35$\,AU. If the grains are small and not blackbody emitters, the emissivity will be higher at shorter wavelengths, thus strengthening this constraint. The central hole of any debris disk is significantly larger than the Keplerian orbit corresponding to the transit signal ($0.051\pm0.004$\,AU). This suggests that the planet and disk are physically separated and unrelated phenomena.

Disks around stars can create ``dips'' in the light curve as vertical structures in the disks periodically occult the host star \citep[e.g.,][]{2014AJ....147...82C}. This behavior has been observed in some stars in Upper Scorpius \citep[e.g.,][]{2016ApJ...816...69A}. However, such dips are usually much deeper and are quasi-periodic or periodic, with the depth changing from dip to dip. The shapes of the dips typically do not resemble a transit; they are irregular and/or have leading or lagging tails \citep{2016ApJ...816...69A}. In contrast, the signals in the lightcurve of \name\ are strictly periodic and transit-shaped, exhibiting no changes over the $\sim1.5$~yr interval between the \ktwo\ and \mearth\ observations. Finally all such ``dipper'' stars exhibit significant excess emission at 12 and 24\,\um\ consistent with full or evolved disks, which is not the case for \name. 

\begin{figure}
	\centering
	\includegraphics[width=0.46\textwidth]{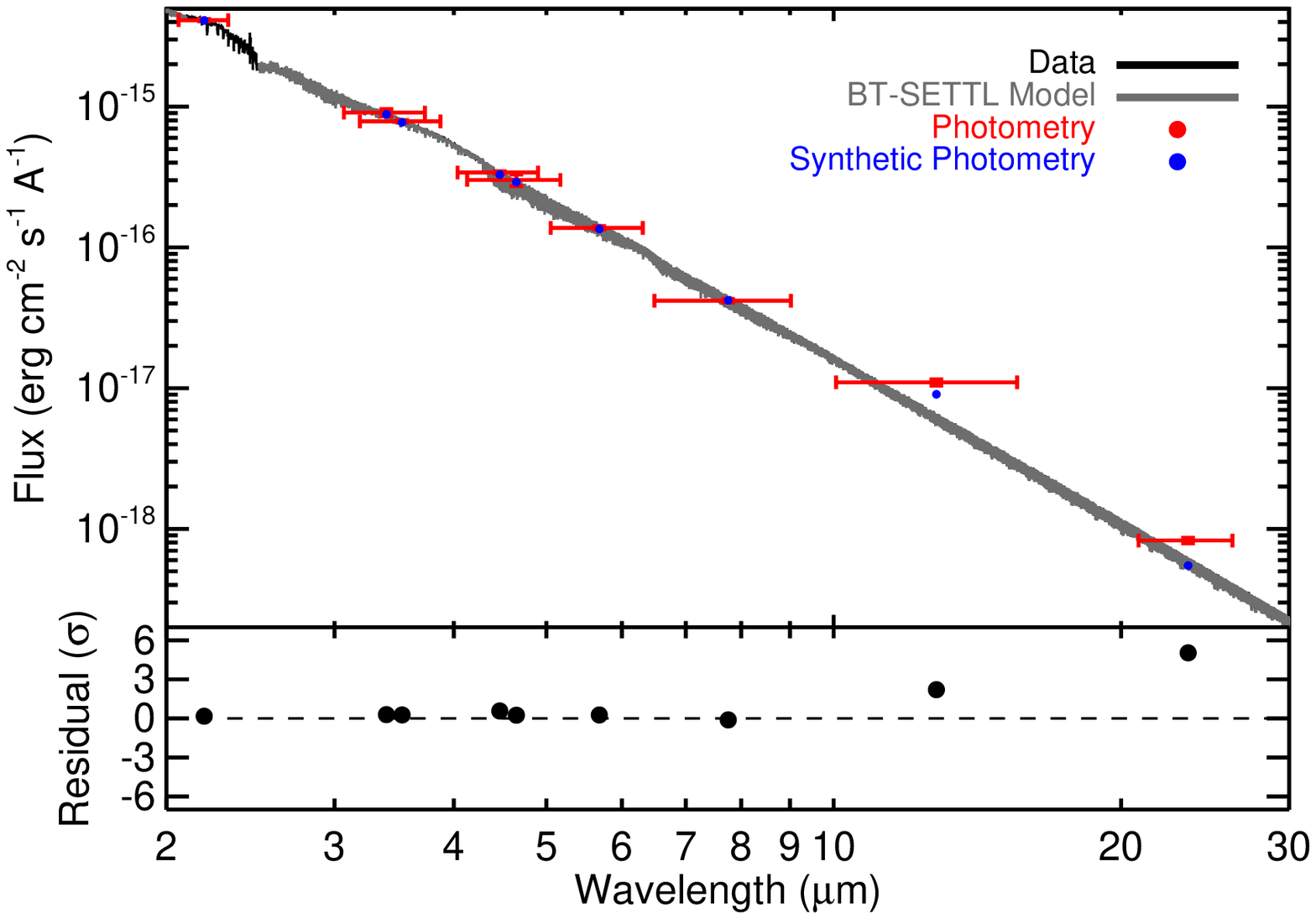} 
	\caption{Spectrum of \name\ following the format of Figure~\ref{fig:spec}, but extended into the infrared and with log scaling on both axes. Only the 24\,\um\ observation by {\it Spitzer}-MPIS shows a statistically significant excess, suggesting the presence of a cold ($<300$\,K) debris disk. }
	\label{fig:IR}
\end{figure}

\section{Summary \& Discussion}\label{sec:discussion} 
Young stellar associations like Upper Scorpius offer a unique view of the properties and behavior of young stars. Planets around these young stars are similarly critical probes into how planets change from formation to maturity \citep[e.g.,][]{2007MNRAS.375...29A, 2012ApJ...756L..33Q,Bowler2015a, David2016}. In this paper we have detailed our follow-up, characterization, and confirmation of \pname, a \prad\ planet orbiting at a period of 5.425~days around a star in the $\sim$11\,Myr \citep{2012ApJ...746..154P} Upper Scorpius OB association. 

In combination, \name's proper motions, radial velocity, and lithium and mid-IR excess in its spectrum, unambiguously indicate that \name\ is a young ($<20$\,Myr) pre-main-sequence star associated with the Upper Scorpius association. We used moderate resolution spectra to revise the reddening, \teff, and \fbol\ determinations. We fit the \ktwo\ and \mearth\ transit photometry, which also yielded a precise stellar density. By interpolating these constraints onto a grid of pre-main sequence models that imitate magnetic effects on the star's internal structure, we derived a precise radius (6-7\%) and mass (16\%) for \name. 

The data argue strongly that the transit signal is planetary in origin. Adaptive optics and radial velocities rule out a background or bound eclipsing binary as the source of the transit signal. Our \mearth\ transit photometry combined with the \ktwo\ photometry rules out stellar variability or a disk mimicking a transit. Further, \pname\ exhibits none of the unusual light curve behavior of PTFO 8-8695b, the candidate whose V-shaped transit signature exhibits time-variable depth and width \citep{2012ApJ...755...42V, 2015ApJ...809...42C, 2015ApJ...812...48Y}. 

Few young ($<20$\,Myr) stars have their masses/radii determined to better than 10\%, with the exception of young eclipsing binaries \citep[e.g., ][]{2015ApJ...807....3K, 2016ApJ...816...21D}. The precision is in part due to our careful measurement of \fbol\ and \teff, and the additional constraint from the transit lightcurve fit, which provides a stellar density accurate to $\lesssim$10\%. This highlights the power of transiting exoplanets to probe stellar astrophysics. While our current method relies on a stellar model, when {\it Gaia} parallaxes become available \citep{2001A&A...369..339P, 2012Ap&SS.341...31D} we can instead use transiting planets to test these models.

Our stellar density assumes that the planet has zero orbital eccentricity. This is expected for a young Neptune-mass planet where recent interactions with the nascent disk have dampened eccentricity \citep[e.g.,][]{2004ApJ...602..388T, 2007A&A...473..329C}. However, there is a dearth of young planetary systems with known eccentricities that would be necessary to confirm this observationally. To test the sensitivity of our results to the assumption of $e = 0$ we reran our stellar isochrone fits, but instead used a uniform density on $\rho_*$. This gave a best-fit stellar density and radius consistent with our earlier determination, but with a factor of two larger errors (see Table~\ref{tab:stellar}, Figure~\ref{fig:density}). Similarly, we derived a consistent stellar radius when using the Stefan-Boltzman relation, which is independent of both the transit-fit density and stellar models. We conclude that the accuracy of our stellar (and therefore planetary) radius is insensitive to this assumption.

Assuming a Neptune-like density, \pname\ will have a RV amplitude of $\sim$20\,\mps. While this is well within detection limits of current radial velocity instruments, it is smaller than the expected spot-induced RV jitter (100-200\,\mps). Moving to the NIR can reduce this noise source, but not eliminate it \citep{2011ApJ...736..123M,Crockett2012}. Because the planet and stellar rotation periods are known from the light curve, it may be possible to fit each separately with sufficient data and baseline. \name\ will be an excellent target for upcoming NIR radial velocity spectrographs \citep[e.g.,][]{2010SPIE.7735E..37Q,2014SPIE.9147E..15A,Kotani2014}.

As with \ktwo-25b, \pname\ is considerably larger than close-in planets found around similar-mass stars by \kepler. Most planets around M dwarfs found by \kepler\ are 1-2.5$R_\earth$ \citep{Morton2014a, Mulders2015, Dressing2015,Gaidos2016a}, while \pname\ is roughly twice this size at \prad. \pname\ is less of a radius outlier than \ktwo-25b, which orbits a 0.3$M_\odot$ star. \name\ has a mass of 0.55$M_\odot$ and larger planets are more common around higher mass hosts. Further, unlike with the nearby, bright, main-sequence, and photometrically well-behaved stars in the Hyades, it is not clear if our survey is sensitive to the more common 1-2$R_\earth$ planets around similar-mass host stars in Upper Scorpius. However, \pname\ fits into an emerging picture that young planets are larger than their older counterparts. \citet{Mann2016} suggested these large radii could be due to the initial heat of formation as well as inflation and escape of the atmosphere under the influence of the young, active host star \citep[e.g.,][]{Rogers:2011lr,Ehrenreich2015}.

The upper limit on \pname's age provided by its $\simeq$11\,Myr stellar host suggests that it either migrated inwards via disk migration or formed in-situ, as planet-star and planet-planet interactions work on much longer timescales \citep{Fabrycky:2007ys, 2008ApJ...678..498N}, and the conditions for Kozai-Lidov evolution only begin after the disk dissipates \citep[e.g.,][]{2016MNRAS.tmp..399M}. This discovery makes it unlikely that such long-term dynamical interactions are responsible for {\it all} close-in planets. However, it is difficult to draw conclusions about the dominant migration or formation mechanism for close-in planets given the sample size and incomplete understanding of our transit-search pipeline's completeness.

Selection effects may be important here. It is possible that \name\ has an atypical history of formation and migration that also made it the easiest (and hence first) planet of this age to be identified and confirmed. A full search of all young clusters and stellar associations surveyed by the \ktwo\ mission, with proper treatment of detection completeness is underway. This, along with improved statistics provided by the TESS and PLATO missions, will provide an estimate of the planet occurrence rate as a function of time. Trends (or a lack of trends) in this occurrence rate could set constraints on planetary migration timescales. 

\acknowledgements
During the final stages of the analysis for this paper we were informed by another team that an independent analysis of this system was about to be submitted (David et al. Nature, accepted). 

We thank the anonymous referee for their thoughtful comments on the paper. The authors thank Ann Marie Cody for her help with the \ktwo\ light curves when initially identifying the transit. We also thank Charlie for his help with the data analysis and proofreading the paper.

AWM was supported through Hubble Fellowship grant 51364 awarded by the Space Telescope Science Institute, which is operated by the Association of Universities for Research in Astronomy, Inc., for NASA, under contract NAS 5-26555. This research was supported by NASA grant NNX11AC33G to EG. K.R.C. acknowledges support provided by the NSF through grant AST-1449476. A.V. is supported by the NSF Graduate Research Fellowship, Grant No. DGE 1144152. 

The MEarth Team gratefully acknowledges funding from the David and Lucille Packard Fellowship for Science and Engineering (awarded to D.C.). This material is based in part upon work supported by the National Science Foundation under grants AST-0807690, AST-1109468, and AST-1004488 (Alan T. Waterman Award). This publication was made possible through the support of a grant from the John Templeton Foundation. The opinions expressed in this publication are those of the authors and do not necessarily reflect the views of the John Templeton Foundation.

The ARCoIRIS observations were conducted at Cerro Tololo Inter-American Observatory, National Optical Astronomy Observatory, which is operated by the Association of Universities for Research in Astronomy (AURA) under a cooperative agreement with the National Science Foundation. This work used the Immersion Grating Infrared Spectrograph (IGRINS) that was developed under a collaboration between the University of Texas at Austin and the Korea Astronomy and Space Science Institute (KASI) with the financial support of the US National Science Foundation under grant ASTR1229522, of the University of Texas at Austin, and of the Korean GMT Project of KASI. The IGRINS pipeline package PLP was developed by Dr. Jae-Joon Lee at Korea Astronomy and Space Science Institute and Professor Soojong Pak's team at Kyung Hee University. SNIFS on the UH 2.2-m telescope is part of the Nearby Supernova Factory project, a scientific collaboration among the Centre de Recherche Astronomique de Lyon, Institut de Physique Nucl\'{e}aire de Lyon, Laboratoire de Physique Nucl\'{e}aire et des Hautes Energies, Lawrence Berkeley National Laboratory, Yale University, University of Bonn, Max Planck Institute for Astrophysics, Tsinghua Center for Astrophysics, and the Centre de Physique des Particules de Marseille. Some of the data presented in this paper were obtained from the Mikulski Archive for Space Telescopes (MAST). STScI is operated by the Association of Universities for Research in Astronomy, Inc., under NASA contract NAS5-26555. Support for MAST for non-HST data is provided by the NASA Office of Space Science via grant NNX09AF08G and by other grants and contracts. This research was made possible through the use of the AAVSO Photometric All-Sky Survey (APASS), funded by the Robert Martin Ayers Sciences Fund. 

The authors wish to recognize and acknowledge the very significant cultural role and reverence that the summit of Maunakea has always had within the indigenous Hawaiian community. We are most fortunate to have the opportunity to conduct observations from this mountain.\\

\facilities{Blanco (ARCoIRIS), UH:2.2m (SNIFS), Keck:II (NIRC2), Smith (IGRINS), Kepler}
\software{\href{http://irtfweb.ifa.hawaii.edu/~spex/}{Spextool} \citep{Cushing2004}, xtellcor \citep{Vacca2003}, \href{https://github.com/igrins/plp}{IGRINS pipeline} \citep{IGRINS_plp}, \href{https://www.dropbox.com/sh/wew08hcqluib8am/AAC6HiCeAVoiDPSe-MXRU2Cda?dl=0}{ARCoIRIS Spextool}, \href{http://astro.uchicago.edu/\~kreidberg/batman/}{batman} \citep{Kreidberg2015}, \href{http://dan.iel.fm/emcee/current/}{emcee} \citep{Foreman-Mackey2013}}

\bibliography{fullbiblio.bib}

\end{document}

%% file: setup.tex
\usepackage{amsmath,amssymb}
\usepackage{epsfig}    
\usepackage{graphicx}    
\usepackage{lineno}
\usepackage{natbib}
\usepackage{bigints}
\usepackage[outdir=./]{epstopdf}

\usepackage[T1]{fontenc}
\pwifjournal\else
  \usepackage{microtype}
\fi

\pwifjournal\else
  \makeatletter
  \renewcommand\plotone[1]{%
    \centering \leavevmode \setlength{\plot@width}{0.95\linewidth}
    \includegraphics[width={\eps@scaling\plot@width}]{#1}%
  }%
  \makeatother
\fi

\makeatletter

\newcommand\@simpfx{http://simbad.u-strasbg.fr/simbad/sim-id?Ident=}

\newcommand\MakeObj[4][\@empty]{
  \pwifjournal%
    \expandafter\newcommand\csname pkgwobj@c@#2\endcsname[1]{\protect\object[#4]{##1}}%
  \else%
    \expandafter\newcommand\csname pkgwobj@c@#2\endcsname[1]{\href{\@simpfx #3}{##1}}%
  \fi%
  \expandafter\newcommand\csname pkgwobj@f#2\endcsname{#4}%
  \ifx\@empty#1%
    \expandafter\newcommand\csname pkgwobj@s#2\endcsname{#4}%
  \else%
    \expandafter\newcommand\csname pkgwobj@s#2\endcsname{#1}%
  \fi}%

\newcommand\MakeTrunc[2]{
  \expandafter\newcommand\csname pkgwobj@t#1\endcsname{#2}}%

\newcommand{\obj}[1]{%
  \expandafter\ifx\csname pkgwobj@c@#1\endcsname\relax%
    \textbf{[unknown object!]}%
  \else%
    \csname pkgwobj@c@#1\endcsname{\csname pkgwobj@s#1\endcsname}%
  \fi}
\newcommand{\objf}[1]{%
  \expandafter\ifx\csname pkgwobj@c@#1\endcsname\relax%
    \textbf{[unknown object!]}%
  \else%
    \csname pkgwobj@c@#1\endcsname{\csname pkgwobj@f#1\endcsname}%
  \fi}
\newcommand{\objt}[1]{%
  \expandafter\ifx\csname pkgwobj@c@#1\endcsname\relax%
    \textbf{[unknown object!]}%
  \else%
    \csname pkgwobj@c@#1\endcsname{\csname pkgwobj@t#1\endcsname}%
  \fi}

\makeatother

\pwifjournal\else
  \usepackage{etoolbox}
  \makeatletter
  \patchcmd{\NAT@citex}
    {\@citea\NAT@hyper@{%
       \NAT@nmfmt{\NAT@nm}%
       \hyper@natlinkbreak{\NAT@aysep\NAT@spacechar}{\@citeb\@extra@b@citeb}%
       \NAT@date}}
    {\@citea\NAT@nmfmt{\NAT@nm}%
     \NAT@aysep\NAT@spacechar\NAT@hyper@{\NAT@date}}{}{}
  \patchcmd{\NAT@citex}
    {\@citea\NAT@hyper@{%
       \NAT@nmfmt{\NAT@nm}%
       \hyper@natlinkbreak{\NAT@spacechar\NAT@@open\if*#1*\else#1\NAT@spacechar\fi}%
         {\@citeb\@extra@b@citeb}%
       \NAT@date}}
    {\@citea\NAT@nmfmt{\NAT@nm}%
     \NAT@spacechar\NAT@@open\if*#1*\else#1\NAT@spacechar\fi\NAT@hyper@{\NAT@date}}
    {}{}
  \makeatother
\fi

\MakeObj{n33370}{NLTT\%2033370}{NLTT~33370\,AB}
\MakeTrunc{n33370}{NLTT~33370}
\MakeObj{tvlm}{TVLM\%20513-46546}{TVLM~513--46546}

\providecommand{\adsurl}[1]{\href{#1}{ADS}}
\newcommand{\mps}{m\,s$^{-1}$}

\newcommand{\vsini}{$v\sin{i_*}$}

\newcommand{\kepler}{{\it Kepler}}

\newcommand{\um}{$\mu$m}
\newcommand{\fbol}{$F_{\mathrm{bol}}$}

\newcommand{\rchisq}{$\chi^2_{\nu}$}

\newcommand\teff{\ensuremath{T_\text{eff}}}

\newcommand\kms{km~s$^{-1}$}

%% file: ms.bbl
\begin{thebibliography}{}
\providecommand\natexlab[1]{#1}
\providecommand\JournalTitle[1]{#1}

\bibitem[{{Ahn} {et~al.}(2012){Ahn}, {Alexandroff}, {Allende Prieto},
  {Anderson}, {Anderton}, {Andrews}, {Aubourg}, {Bailey}, {Balbinot}, {Barnes},
  \& et~al.}]{Ahn:2012kx}
{Ahn}, C.~P., {Alexandroff}, R., {Allende Prieto}, C., {et~al.} 2012,
  \href{http://dx.doi.org/10.1088/0067-0049/203/2/21}{\JournalTitle{\apjs},
  203, 21}

\bibitem[{{Aigrain} {et~al.}(2007){Aigrain}, {Hodgkin}, {Irwin}, {Hebb},
  {Irwin}, {Favata}, {Moraux}, \& {Pont}}]{2007MNRAS.375...29A}
{Aigrain}, S., {Hodgkin}, S., {Irwin}, J., {et~al.} 2007,
  \href{http://dx.doi.org/10.1111/j.1365-2966.2006.11303.x}{\JournalTitle{\mnras},
  375, 29}

\bibitem[{{Aldering} {et~al.}(2002){Aldering}, {Adam}, {Antilogus}, {Astier},
  {Bacon}, {Bongard}, {Bonnaud}, {Copin}, {Hardin}, {Henault}, {Howell},
  {Lemonnier}, {Levy}, {Loken}, {Nugent}, {Pain}, {Pecontal}, {Pecontal},
  {Perlmutter}, {Quimby}, {Schahmaneche}, {Smadja}, \&
  {Wood-Vasey}}]{Aldering2002}
{Aldering}, G., {Adam}, G., {Antilogus}, P., {et~al.} 2002,
  \href{http://dx.doi.org/10.1117/12.458107}{in Society of Photo-Optical
  Instrumentation Engineers (SPIE) Conference Series, Vol. 4836, Survey and
  Other Telescope Technologies and Discoveries, ed. J.~A. {Tyson} \&
  S.~{Wolff}}, 61

\bibitem[{{Allard} {et~al.}(2011){Allard}, {Homeier}, \&
  {Freytag}}]{Allard2011}
{Allard}, F., {Homeier}, D., \& {Freytag}, B. 2011, in Astronomical Society of
  the Pacific Conference Series, Vol. 448, 16th Cambridge Workshop on Cool
  Stars, Stellar Systems, and the Sun, ed. C.~{Johns-Krull}, M.~K. {Browning},
  \& A.~A. {West}, 91

\bibitem[{{Allard} {et~al.}(2012){Allard}, {Homeier}, {Freytag}, \&
  {Sharp}}]{Allard:2012fk}
{Allard}, F., {Homeier}, D., {Freytag}, B., \& {Sharp}, C.~M. 2012,
  \href{http://dx.doi.org/10.1051/eas/1257001}{in EAS Publications Series,
  Vol.~57, EAS Publications Series}, 3

\bibitem[{{Ansdell} {et~al.}(2016){Ansdell}, {Gaidos}, {Rappaport}, {Jacobs},
  {LaCourse}, {Jek}, {Mann}, {Wyatt}, {Kennedy}, {Williams}, \&
  {Boyajian}}]{2016ApJ...816...69A}
{Ansdell}, M., {Gaidos}, E., {Rappaport}, S.~A., {et~al.} 2016,
  \href{http://dx.doi.org/10.3847/0004-637X/816/2/69}{\JournalTitle{\apj}, 816,
  69}

\bibitem[{{Artigau} {et~al.}(2014){Artigau}, {Kouach}, {Donati}, {Doyon},
  {Delfosse}, {Baratchart}, {Lacombe}, {Moutou}, {Rabou}, {Par{\`e}s},
  {Micheau}, {Thibault}, {Reshetov}, {Dubois}, {Hernandez}, {Vall{\'e}e},
  {Wang}, {Dolon}, {Pepe}, {Bouchy}, {Striebig}, {H{\'e}nault}, {Loop},
  {Saddlemyer}, {Barrick}, {Vermeulen}, {Dupieux}, {H{\'e}brard}, {Boisse},
  {Martioli}, {Alencar}, {do Nascimento}, \& {Figueira}}]{2014SPIE.9147E..15A}
{Artigau}, {\'E}., {Kouach}, D., {Donati}, J.-F., {et~al.} 2014,
  \href{http://dx.doi.org/10.1117/12.2055663}{in Society of Photo-Optical
  Instrumentation Engineers (SPIE) Conference Series, Vol. 9147, Society of
  Photo-Optical Instrumentation Engineers (SPIE) Conference Series}

\bibitem[{{Becker} {et~al.}(2015){Becker}, {Vanderburg}, {Adams}, {Rappaport},
  \& {Schwengeler}}]{Becker2015}
{Becker}, J.~C., {Vanderburg}, A., {Adams}, F.~C., {Rappaport}, S.~A., \&
  {Schwengeler}, H.~M. 2015,
  \href{http://dx.doi.org/10.1088/2041-8205/812/2/L18}{\JournalTitle{\apjl},
  812, L18}

\bibitem[{{Berta} {et~al.}(2012){Berta}, {Charbonneau}, {D{\'e}sert},
  {Miller-Ricci Kempton}, {McCullough}, {Burke}, {Fortney}, {Irwin}, {Nutzman},
  \& {Homeier}}]{2012ApJ...747...35B}
{Berta}, Z.~K., {Charbonneau}, D., {D{\'e}sert}, J.-M., {et~al.} 2012,
  \href{http://dx.doi.org/10.1088/0004-637X/747/1/35}{\JournalTitle{\apj}, 747,
  35}

\bibitem[{{Bowler} {et~al.}(2015{\natexlab{a}}){Bowler}, {Liu}, {Shkolnik}, \&
  {Tamura}}]{Bowler2015a}
{Bowler}, B.~P., {Liu}, M.~C., {Shkolnik}, E.~L., \& {Tamura}, M.
  2015{\natexlab{a}},
  \href{http://dx.doi.org/10.1088/0067-0049/216/1/7}{\JournalTitle{\apjs}, 216,
  7}

\bibitem[{{Bowler} {et~al.}(2015{\natexlab{b}}){Bowler}, {Shkolnik}, {Liu},
  {Schlieder}, {Mann}, {Dupuy}, {Hinkley}, {Crepp}, {Johnson}, {Howard},
  {Flagg}, {Weinberger}, {Aller}, {Allers}, {Best}, {Kotson}, {Montet},
  {Herczeg}, {Baranec}, {Riddle}, {Law}, {Nielsen}, {Wahhaj}, {Biller}, \&
  {Hayward}}]{Bowler2015b}
{Bowler}, B.~P., {Shkolnik}, E.~L., {Liu}, M.~C., {et~al.} 2015{\natexlab{b}},
  \href{http://dx.doi.org/10.1088/0004-637X/806/1/62}{\JournalTitle{\apj}, 806,
  62}

\bibitem[{{Boyajian} {et~al.}(2012){Boyajian}, {von Braun}, {van Belle},
  {McAlister}, {ten Brummelaar}, {Kane}, {Muirhead}, {Jones}, {White},
  {Schaefer}, {Ciardi}, {Henry}, {L{\'o}pez-Morales}, {Ridgway}, {Gies}, {Jao},
  {Rojas-Ayala}, {Parks}, {Sturmann}, {Sturmann}, {Turner}, {Farrington},
  {Goldfinger}, \& {Berger}}]{Boyajian2012}
{Boyajian}, T.~S., {von Braun}, K., {van Belle}, G., {et~al.} 2012,
  \href{http://dx.doi.org/10.1088/0004-637X/757/2/112}{\JournalTitle{\apj},
  757, 112}

\bibitem[{{Brandt} \& {Huang}(2015)}]{Brandt2015}
{Brandt}, T.~D., \& {Huang}, C.~X. 2015,
  \href{http://dx.doi.org/10.1088/0004-637X/807/1/24}{\JournalTitle{\apj}, 807,
  24}

\bibitem[{{Bubar} {et~al.}(2011){Bubar}, {Schaeuble}, {King}, {Mamajek}, \&
  {Stauffer}}]{2011AJ....142..180B}
{Bubar}, E.~J., {Schaeuble}, M., {King}, J.~R., {Mamajek}, E.~E., \&
  {Stauffer}, J.~R. 2011,
  \href{http://dx.doi.org/10.1088/0004-6256/142/6/180}{\JournalTitle{\aj}, 142,
  180}

\bibitem[{{Cardelli} {et~al.}(1989){Cardelli}, {Clayton}, \&
  {Mathis}}]{1989ApJ...345..245C}
{Cardelli}, J.~A., {Clayton}, G.~C., \& {Mathis}, J.~S. 1989,
  \href{http://dx.doi.org/10.1086/167900}{\JournalTitle{\apj}, 345, 245}

\bibitem[{{Carpenter} {et~al.}(2009){Carpenter}, {Bouwman}, {Mamajek}, {Meyer},
  {Hillenbrand}, {Backman}, {Henning}, {Hines}, {Hollenbach}, {Kim},
  {Moro-Martin}, {Pascucci}, {Silverstone}, {Stauffer}, \&
  {Wolf}}]{2009ApJS..181..197C}
{Carpenter}, J.~M., {Bouwman}, J., {Mamajek}, E.~E., {et~al.} 2009,
  \href{http://dx.doi.org/10.1088/0067-0049/181/1/197}{\JournalTitle{\apjs},
  181, 197}

\bibitem[{{Chatterjee} {et~al.}(2008){Chatterjee}, {Ford}, {Matsumura}, \&
  {Rasio}}]{2008ApJ...686..580C}
{Chatterjee}, S., {Ford}, E.~B., {Matsumura}, S., \& {Rasio}, F.~A. 2008,
  \href{http://dx.doi.org/10.1086/590227}{\JournalTitle{\apj}, 686, 580}

\bibitem[{{Chen} {et~al.}(2014){Chen}, {Girardi}, {Bressan}, {Marigo},
  {Barbieri}, \& {Kong}}]{Chen2014}
{Chen}, Y., {Girardi}, L., {Bressan}, A., {et~al.} 2014,
  \href{http://dx.doi.org/10.1093/mnras/stu1605}{\JournalTitle{\mnras}, 444,
  2525}

\bibitem[{{Chiang} \& {Laughlin}(2013)}]{2013MNRAS.431.3444C}
{Chiang}, E., \& {Laughlin}, G. 2013,
  \href{http://dx.doi.org/10.1093/mnras/stt424}{\JournalTitle{\mnras}, 431,
  3444}

\bibitem[{{Ciardi} {et~al.}(2015){Ciardi}, {van Eyken}, {Barnes}, {Beichman},
  {Carey}, {Crockett}, {Eastman}, {Johns-Krull}, {Howell}, {Kane}, {.~Mclane},
  {Plavchan}, {Prato}, {Stauffer}, {van Belle}, \& {von
  Braun}}]{2015ApJ...809...42C}
{Ciardi}, D.~R., {van Eyken}, J.~C., {Barnes}, J.~W., {et~al.} 2015,
  \href{http://dx.doi.org/10.1088/0004-637X/809/1/42}{\JournalTitle{\apj}, 809,
  42}

\bibitem[{{Cody} {et~al.}(2014){Cody}, {Stauffer}, {Baglin}, {Micela},
  {Rebull}, {Flaccomio}, {Morales-Calder{\'o}n}, {Aigrain}, {Bouvier},
  {Hillenbrand}, {Gutermuth}, {Song}, {Turner}, {Alencar}, {Zwintz},
  {Plavchan}, {Carpenter}, {Findeisen}, {Carey}, {Terebey}, {Hartmann},
  {Calvet}, {Teixeira}, {Vrba}, {Wolk}, {Covey}, {Poppenhaeger}, {G{\"u}nther},
  {Forbrich}, {Whitney}, {Affer}, {Herbst}, {Hora}, {Barrado}, {Holtzman},
  {Marchis}, {Wood}, {Medeiros Guimar{\~a}es}, {Lillo Box}, {Gillen},
  {McQuillan}, {Espaillat}, {Allen}, {D'Alessio}, \&
  {Favata}}]{2014AJ....147...82C}
{Cody}, A.~M., {Stauffer}, J., {Baglin}, A., {et~al.} 2014,
  \href{http://dx.doi.org/10.1088/0004-6256/147/4/82}{\JournalTitle{\aj}, 147,
  82}

\bibitem[{{Cohen} {et~al.}(2003){Cohen}, {Wheaton}, \&
  {Megeath}}]{2003AJ....126.1090C}
{Cohen}, M., {Wheaton}, W.~A., \& {Megeath}, S.~T. 2003,
  \href{http://dx.doi.org/10.1086/376474}{\JournalTitle{\aj}, 126, 1090}

\bibitem[{{Cresswell} {et~al.}(2007){Cresswell}, {Dirksen}, {Kley}, \&
  {Nelson}}]{2007A&A...473..329C}
{Cresswell}, P., {Dirksen}, G., {Kley}, W., \& {Nelson}, R.~P. 2007,
  \href{http://dx.doi.org/10.1051/0004-6361:20077666}{\JournalTitle{\aap}, 473,
  329}

\bibitem[{{Crockett} {et~al.}(2012){Crockett}, {Mahmud}, {Prato},
  {Johns-Krull}, {Jaffe}, {Hartigan}, \& {Beichman}}]{Crockett2012}
{Crockett}, C.~J., {Mahmud}, N.~I., {Prato}, L., {et~al.} 2012,
  \href{http://dx.doi.org/10.1088/0004-637X/761/2/164}{\JournalTitle{\apj},
  761, 164}

\bibitem[{{Cushing} {et~al.}(2004){Cushing}, {Vacca}, \&
  {Rayner}}]{Cushing2004}
{Cushing}, M.~C., {Vacca}, W.~D., \& {Rayner}, J.~T. 2004,
  \href{http://dx.doi.org/10.1086/382907}{\JournalTitle{\pasp}, 116, 362}

\bibitem[{{Dahm}(2015)}]{2015ApJ...813..108D}
{Dahm}, S.~E. 2015,
  \href{http://dx.doi.org/10.1088/0004-637X/813/2/108}{\JournalTitle{\apj},
  813, 108}

\bibitem[{{David} {et~al.}(2016{\natexlab{a}}){David}, {Hillenbrand}, {Cody},
  {Carpenter}, \& {Howard}}]{2016ApJ...816...21D}
{David}, T.~J., {Hillenbrand}, L.~A., {Cody}, A.~M., {Carpenter}, J.~M., \&
  {Howard}, A.~W. 2016{\natexlab{a}},
  \href{http://dx.doi.org/10.3847/0004-637X/816/1/21}{\JournalTitle{\apj}, 816,
  21}

\bibitem[{{David} {et~al.}(2016{\natexlab{b}}){David}, {Conroy}, {Hillenbrand},
  {Stassun}, {Stauffer}, {Rebull}, {Cody}, {Isaacson}, {Howard}, \&
  {Aigrain}}]{David2016}
{David}, T.~J., {Conroy}, K.~E., {Hillenbrand}, L.~A., {et~al.}
  2016{\natexlab{b}},
  \href{http://dx.doi.org/10.3847/0004-6256/151/5/112}{\JournalTitle{\aj}, 151,
  112}

\bibitem[{{de Bruijne}(2012)}]{2012Ap&SS.341...31D}
{de Bruijne}, J.~H.~J. 2012,
  \href{http://dx.doi.org/10.1007/s10509-012-1019-4}{\JournalTitle{\apss}, 341,
  31}

\bibitem[{{de Zeeuw} {et~al.}(1999){de Zeeuw}, {Hoogerwerf}, {de Bruijne},
  {Brown}, \& {Blaauw}}]{1999AJ....117..354D}
{de Zeeuw}, P.~T., {Hoogerwerf}, R., {de Bruijne}, J.~H.~J., {Brown}, A.~G.~A.,
  \& {Blaauw}, A. 1999,
  \href{http://dx.doi.org/10.1086/300682}{\JournalTitle{\aj}, 117, 354}

\bibitem[{{Dittmann} {et~al.}(2016){Dittmann}, {Irwin}, {Charbonneau}, \&
  {Newton}}]{2016ApJ...818..153D}
{Dittmann}, J.~A., {Irwin}, J.~M., {Charbonneau}, D., \& {Newton}, E.~R. 2016,
  \href{http://dx.doi.org/10.3847/0004-637X/818/2/153}{\JournalTitle{\apj},
  818, 153}

\bibitem[{{Dotter} {et~al.}(2008){Dotter}, {Chaboyer}, {Jevremovi{\'c}},
  {Kostov}, {Baron}, \& {Ferguson}}]{Dotter2008}
{Dotter}, A., {Chaboyer}, B., {Jevremovi{\'c}}, D., {et~al.} 2008,
  \href{http://dx.doi.org/10.1086/589654}{\JournalTitle{\apjs}, 178, 89}

\bibitem[{{Dressing} \& {Charbonneau}(2015)}]{Dressing2015}
{Dressing}, C.~D., \& {Charbonneau}, D. 2015,
  \href{http://dx.doi.org/10.1088/0004-637X/807/1/45}{\JournalTitle{\apj}, 807,
  45}

\bibitem[{{Ehrenreich} {et~al.}(2015){Ehrenreich}, {Bourrier}, {Wheatley}, {Des
  Etangs}, {H{\'e}brard}, {Udry}, {Bonfils}, {Delfosse}, {D{\'e}sert}, {Sing},
  \& {Vidal-Madjar}}]{Ehrenreich2015}
{Ehrenreich}, D., {Bourrier}, V., {Wheatley}, P.~J., {et~al.} 2015,
  \href{http://dx.doi.org/10.1038/nature14501}{\JournalTitle{\nat}, 522, 459}

\bibitem[{{Fabrycky} \& {Tremaine}(2007)}]{Fabrycky:2007ys}
{Fabrycky}, D., \& {Tremaine}, S. 2007,
  \href{http://dx.doi.org/10.1086/521702}{\JournalTitle{\apj}, 669, 1298}

\bibitem[{{Feiden}(2016)}]{Feiden2016}
{Feiden}, G.~A. 2016, \JournalTitle{ArXiv e-prints},
  \href{http://arxiv.org/abs/1604.08036}{{\sffamily arXiv:1604.08036
  [astro-ph.SR]}}

\bibitem[{{Feiden} \& {Chaboyer}(2012)}]{Feiden2012b}
{Feiden}, G.~A., \& {Chaboyer}, B. 2012,
  \href{http://dx.doi.org/10.1088/0004-637X/761/1/30}{\JournalTitle{\apj}, 761,
  30}

\bibitem[{{Feiden} \& {Chaboyer}(2013)}]{Feiden2013}
---. 2013,
  \href{http://dx.doi.org/10.1088/0004-637X/779/2/183}{\JournalTitle{\apj},
  779, 183}

\bibitem[{{Ford} \& {Rasio}(2006)}]{2006ApJ...638L..45F}
{Ford}, E.~B., \& {Rasio}, F.~A. 2006,
  \href{http://dx.doi.org/10.1086/500734}{\JournalTitle{\apjl}, 638, L45}

\bibitem[{{Foreman-Mackey} {et~al.}(2013){Foreman-Mackey}, {Hogg}, {Lang}, \&
  {Goodman}}]{Foreman-Mackey2013}
{Foreman-Mackey}, D., {Hogg}, D.~W., {Lang}, D., \& {Goodman}, J. 2013,
  \href{http://dx.doi.org/10.1086/670067}{\JournalTitle{\pasp}, 125, 306}

\bibitem[{{Fressin} {et~al.}(2013){Fressin}, {Torres}, {Charbonneau}, {Bryson},
  {Christiansen}, {Dressing}, {Jenkins}, {Walkowicz}, \&
  {Batalha}}]{Fressin:2013qy}
{Fressin}, F., {Torres}, G., {Charbonneau}, D., {et~al.} 2013,
  \href{http://dx.doi.org/10.1088/0004-637X/766/2/81}{\JournalTitle{\apj}, 766,
  81}

\bibitem[{{Gaidos} {et~al.}(2016{\natexlab{a}}){Gaidos}, {Mann}, \&
  {Ansdell}}]{Gaidos2016}
{Gaidos}, E., {Mann}, A.~W., \& {Ansdell}, M. 2016{\natexlab{a}},
  \href{http://dx.doi.org/10.3847/0004-637X/817/1/50}{\JournalTitle{\apj}, 817,
  50}

\bibitem[{{Gaidos} {et~al.}(2016{\natexlab{b}}){Gaidos}, {Mann}, {Kraus}, \&
  {Ireland}}]{Gaidos2016a}
{Gaidos}, E., {Mann}, A.~W., {Kraus}, A.~L., \& {Ireland}, M.
  2016{\natexlab{b}},
  \href{http://dx.doi.org/10.1093/mnras/stw097}{\JournalTitle{\mnras}, 457,
  2877}

\bibitem[{{Gaidos} {et~al.}(2014){Gaidos}, {Mann}, {L{\'e}pine}, {Buccino},
  {James}, {Ansdell}, {Petrucci}, {Mauas}, \& {Hilton}}]{Gaidos2014}
{Gaidos}, E., {Mann}, A.~W., {L{\'e}pine}, S., {et~al.} 2014,
  \href{http://dx.doi.org/10.1093/mnras/stu1313}{\JournalTitle{\mnras}, 443,
  2561}

\bibitem[{{Gray}(1992)}]{Gray1992}
{Gray}, D.~F. 1992, {The observation and analysis of stellar photospheres.},
  Vol.~10 (Camb.~Astrophys.~Ser)

\bibitem[{{Henden} {et~al.}(2012){Henden}, {Levine}, {Terrell}, {Smith}, \&
  {Welch}}]{Henden:2012fk}
{Henden}, A.~A., {Levine}, S.~E., {Terrell}, D., {Smith}, T.~C., \& {Welch}, D.
  2012, \JournalTitle{Journal of the American Association of Variable Star
  Observers (JAAVSO)}, 40, 430

\bibitem[{{Herczeg} \& {Hillenbrand}(2014)}]{2014ApJ...786...97H}
{Herczeg}, G.~J., \& {Hillenbrand}, L.~A. 2014,
  \href{http://dx.doi.org/10.1088/0004-637X/786/2/97}{\JournalTitle{\apj}, 786,
  97}

\bibitem[{{Howard} {et~al.}(2010){Howard}, {Marcy}, {Johnson}, {Fischer},
  {Wright}, {Isaacson}, {Valenti}, {Anderson}, {Lin}, \&
  {Ida}}]{2010Sci...330..653H}
{Howard}, A.~W., {Marcy}, G.~W., {Johnson}, J.~A., {et~al.} 2010,
  \href{http://dx.doi.org/10.1126/science.1194854}{\JournalTitle{Science}, 330,
  653}

\bibitem[{{Howell} {et~al.}(2014){Howell}, {Sobeck}, {Haas}, {Still},
  {Barclay}, {Mullally}, {Troeltzsch}, {Aigrain}, {Bryson}, {Caldwell},
  {Chaplin}, {Cochran}, {Huber}, {Marcy}, {Miglio}, {Najita}, {Smith},
  {Twicken}, \& {Fortney}}]{Howell2014}
{Howell}, S.~B., {Sobeck}, C., {Haas}, M., {et~al.} 2014,
  \href{http://dx.doi.org/10.1086/676406}{\JournalTitle{\pasp}, 126, 398}

\bibitem[{{Hubeny} \& {Lanz}(2011)}]{2011ascl.soft09022H}
{Hubeny}, I., \& {Lanz}, T. 2011, {Synspec: General Spectrum Synthesis
  Program}, Astrophysics Source Code Library,
  \href{http://arxiv.org/abs/1109.022}{{\sffamily ascl:1109.022}}

\bibitem[{{Husser} {et~al.}(2013){Husser}, {Wende-von Berg}, {Dreizler},
  {Homeier}, {Reiners}, {Barman}, \& {Hauschildt}}]{2013A&A...553A...6H}
{Husser}, T.-O., {Wende-von Berg}, S., {Dreizler}, S., {et~al.} 2013,
  \href{http://dx.doi.org/10.1051/0004-6361/201219058}{\JournalTitle{\aap},
  553, A6}

\bibitem[{{Ida} \& {Lin}(2008)}]{2008ApJ...673..487I}
{Ida}, S., \& {Lin}, D.~N.~C. 2008,
  \href{http://dx.doi.org/10.1086/523754}{\JournalTitle{\apj}, 673, 487}

\bibitem[{{Irwin} {et~al.}(2007){Irwin}, {Irwin}, {Aigrain}, {Hodgkin}, {Hebb},
  \& {Moraux}}]{2007MNRAS.375.1449I}
{Irwin}, J., {Irwin}, M., {Aigrain}, S., {et~al.} 2007,
  \href{http://dx.doi.org/10.1111/j.1365-2966.2006.11408.x}{\JournalTitle{\mnras},
  375, 1449}

\bibitem[{{Irwin} {et~al.}(2015){Irwin}, {Berta-Thompson}, {Charbonneau},
  {Dittmann}, {Falco}, {Newton}, \& {Nutzman}}]{2015csss...18..767I}
{Irwin}, J.~M., {Berta-Thompson}, Z.~K., {Charbonneau}, D., {et~al.} 2015, in
  Cambridge Workshop on Cool Stars, Stellar Systems, and the Sun, Vol.~18, 18th
  Cambridge Workshop on Cool Stars, Stellar Systems, and the Sun, ed. G.~T.
  {van Belle} \& H.~C. {Harris}, 767

\bibitem[{{Janes}(1996)}]{1996JGR...10114853J}
{Janes}, K. 1996,
  \href{http://dx.doi.org/10.1029/96JE00833}{\JournalTitle{\jgr}, 101, 14853}

\bibitem[{{Jarrett} {et~al.}(2011){Jarrett}, {Cohen}, {Masci}, {Wright},
  {Stern}, {Benford}, {Blain}, {Carey}, {Cutri}, {Eisenhardt}, {Lonsdale},
  {Mainzer}, {Marsh}, {Padgett}, {Petty}, {Ressler}, {Skrutskie}, {Stanford},
  {Surace}, {Tsai}, {Wheelock}, \& {Yan}}]{Jarrett2011}
{Jarrett}, T.~H., {Cohen}, M., {Masci}, F., {et~al.} 2011,
  \href{http://dx.doi.org/10.1088/0004-637X/735/2/112}{\JournalTitle{\apj},
  735, 112}

\bibitem[{{Kenworthy} {et~al.}(2015){Kenworthy}, {Lacour}, {Kraus}, {Triaud},
  {Mamajek}, {Scott}, {S{\'e}gransan}, {Ireland}, {Hambsch}, {Reichart},
  {Haislip}, {LaCluyze}, {Moore}, \& {Frank}}]{2015MNRAS.446..411K}
{Kenworthy}, M.~A., {Lacour}, S., {Kraus}, A., {et~al.} 2015,
  \href{http://dx.doi.org/10.1093/mnras/stu2067}{\JournalTitle{\mnras}, 446,
  411}

\bibitem[{{Kenyon} {et~al.}(2008){Kenyon}, {G{\'o}mez}, \&
  {Whitney}}]{Kenyon2008}
{Kenyon}, S.~J., {G{\'o}mez}, M., \& {Whitney}, B.~A. 2008, {Low Mass Star
  Formation in the Taurus-Auriga Clouds}, ed. B.~{Reipurth}, Vol.~1
  (Astronomical Society of the Pacific), 405

\bibitem[{{Kipping}(2010)}]{Kipping:2010lr}
{Kipping}, D.~M. 2010,
  \href{http://dx.doi.org/10.1111/j.1365-2966.2010.17242.x}{\JournalTitle{\mnras},
  408, 1758}

\bibitem[{{Kipping}(2013)}]{Kipping2013}
---. 2013,
  \href{http://dx.doi.org/10.1093/mnras/stt1435}{\JournalTitle{\mnras}, 435,
  2152}

\bibitem[{{Kotani} {et~al.}(2014){Kotani}, {Tamura}, {Suto}, {Nishikawa},
  {Sato}, {Aoki}, {Usuda}, {Kurokawa}, {Kashiwagi}, {Nishiyama}, {Ikeda},
  {Hall}, {Hodapp}, {Hashimoto}, {Morino}, {Okuyama}, {Tanaka}, {Suzuki},
  {Inoue}, {Kwon}, {Suenaga}, {Oh}, {Baba}, {Narita}, {Kokubo}, {Hayano},
  {Izumiura}, {Kambe}, {Kudo}, {Kusakabe}, {Ikoma}, {Hori}, {Omiya}, {Genda},
  {Fukui}, {Fujii}, {Guyon}, {Harakawa}, {Hayashi}, {Hidai}, {Hirano},
  {Kuzuhara}, {Machida}, {Matsuo}, {Nagata}, {Onuki}, {Ogihara}, {Takami},
  {Takato}, {Takahashi}, {Tachinami}, {Terada}, {Kawahara}, \&
  {Yamamuro}}]{Kotani2014}
{Kotani}, T., {Tamura}, M., {Suto}, H., {et~al.} 2014,
  \href{http://dx.doi.org/10.1117/12.2055075}{in Society of Photo-Optical
  Instrumentation Engineers (SPIE) Conference Series, Vol. 9147, Society of
  Photo-Optical Instrumentation Engineers (SPIE) Conference Series}, 14

\bibitem[{{Kov{\'a}cs} {et~al.}(2002){Kov{\'a}cs}, {Zucker}, \&
  {Mazeh}}]{Kovacs2002}
{Kov{\'a}cs}, G., {Zucker}, S., \& {Mazeh}, T. 2002,
  \href{http://dx.doi.org/10.1051/0004-6361:20020802}{\JournalTitle{\aap}, 391,
  369}

\bibitem[{{Kraus} {et~al.}(2015){Kraus}, {Cody}, {Covey}, {Rizzuto}, {Mann}, \&
  {Ireland}}]{2015ApJ...807....3K}
{Kraus}, A.~L., {Cody}, A.~M., {Covey}, K.~R., {et~al.} 2015,
  \href{http://dx.doi.org/10.1088/0004-637X/807/1/3}{\JournalTitle{\apj}, 807,
  3}

\bibitem[{{Kraus} \& {Hillenbrand}(2008)}]{Kraus2008b}
{Kraus}, A.~L., \& {Hillenbrand}, L.~A. 2008,
  \href{http://dx.doi.org/10.1086/593012}{\JournalTitle{\apjl}, 686, L111}

\bibitem[{{Kraus} {et~al.}(2008){Kraus}, {Ireland}, {Martinache}, \&
  {Lloyd}}]{Kraus2008}
{Kraus}, A.~L., {Ireland}, M.~J., {Martinache}, F., \& {Lloyd}, J.~P. 2008,
  \href{http://dx.doi.org/10.1086/587435}{\JournalTitle{\apj}, 679, 762}

\bibitem[{{Kreidberg}(2015)}]{Kreidberg2015}
{Kreidberg}, L. 2015,
  \href{http://dx.doi.org/10.1086/683602}{\JournalTitle{\pasp}, 127, 1161}

\bibitem[{{Lantz} {et~al.}(2004){Lantz}, {Aldering}, {Antilogus}, {Bonnaud},
  {Capoani}, {Castera}, {Copin}, {Dubet}, {Gangler}, {Henault}, {Lemonnier},
  {Pain}, {Pecontal}, {Pecontal}, \& {Smadja}}]{Lantz2004}
{Lantz}, B., {Aldering}, G., {Antilogus}, P., {et~al.} 2004,
  \href{http://dx.doi.org/10.1117/12.512493}{in Society of Photo-Optical
  Instrumentation Engineers (SPIE) Conference Series, Vol. 5249, Optical Design
  and Engineering, ed. L.~{Mazuray}, P.~J. {Rogers}, \& R.~{Wartmann}}, 146

\bibitem[{{Le Bouquin} \& {Absil}(2012)}]{2012A&A...541A..89L}
{Le Bouquin}, J.-B., \& {Absil}, O. 2012,
  \href{http://dx.doi.org/10.1051/0004-6361/201117891}{\JournalTitle{\aap},
  541, A89}

\bibitem[{Lee(2015)}]{IGRINS_plp}
Lee, J.-J. 2015, plp: Version 2.0

\bibitem[{{Lubow} \& {Ida}(2010)}]{2010arXiv1004.4137L}
{Lubow}, S.~H., \& {Ida}, S. 2010, \JournalTitle{ArXiv e-prints},
  \href{http://arxiv.org/abs/1004.4137}{{\sffamily arXiv:1004.4137
  [astro-ph.EP]}}

\bibitem[{{Luhman} \& {Mamajek}(2012)}]{2012ApJ...758...31L}
{Luhman}, K.~L., \& {Mamajek}, E.~E. 2012,
  \href{http://dx.doi.org/10.1088/0004-637X/758/1/31}{\JournalTitle{\apj}, 758,
  31}

\bibitem[{{Mahmud} {et~al.}(2011){Mahmud}, {Crockett}, {Johns-Krull}, {Prato},
  {Hartigan}, {Jaffe}, \& {Beichman}}]{2011ApJ...736..123M}
{Mahmud}, N.~I., {Crockett}, C.~J., {Johns-Krull}, C.~M., {et~al.} 2011,
  \href{http://dx.doi.org/10.1088/0004-637X/736/2/123}{\JournalTitle{\apj},
  736, 123}

\bibitem[{{Mamajek} {et~al.}(2013){Mamajek}, {Pecaut}, {Nguyen}, \&
  {Bubar}}]{2013prpl.conf1K086M}
{Mamajek}, E.~E., {Pecaut}, M.~J., {Nguyen}, D.~C., \& {Bubar}, E.~J. 2013, in
  Protostars and Planets VI Posters

\bibitem[{{Mamajek} {et~al.}(2012){Mamajek}, {Quillen}, {Pecaut}, {Moolekamp},
  {Scott}, {Kenworthy}, {Collier Cameron}, \& {Parley}}]{2012AJ....143...72M}
{Mamajek}, E.~E., {Quillen}, A.~C., {Pecaut}, M.~J., {et~al.} 2012,
  \href{http://dx.doi.org/10.1088/0004-6256/143/3/72}{\JournalTitle{\aj}, 143,
  72}

\bibitem[{{Mandel} \& {Agol}(2002)}]{MandelAgol2002}
{Mandel}, K., \& {Agol}, E. 2002,
  \href{http://dx.doi.org/10.1086/345520}{\JournalTitle{\apjl}, 580, L171}

\bibitem[{{Mann} {et~al.}(2015){Mann}, {Feiden}, {Gaidos}, {Boyajian}, \& {von
  Braun}}]{Mann2015b}
{Mann}, A.~W., {Feiden}, G.~A., {Gaidos}, E., {Boyajian}, T., \& {von Braun},
  K. 2015,
  \href{http://dx.doi.org/10.1088/0004-637X/804/1/64}{\JournalTitle{\apj}, 804,
  64}

\bibitem[{{Mann} {et~al.}(2013){Mann}, {Gaidos}, \& {Ansdell}}]{Mann2013}
{Mann}, A.~W., {Gaidos}, E., \& {Ansdell}, M. 2013,
  \href{http://dx.doi.org/10.1088/0004-637X/779/2/188}{\JournalTitle{\apj},
  779, 188}

\bibitem[{{Mann} \& {von Braun}(2015)}]{Mann2015a}
{Mann}, A.~W., \& {von Braun}, K. 2015,
  \href{http://dx.doi.org/10.1086/680012}{\JournalTitle{\pasp}, 127, 102}

\bibitem[{{Mann} {et~al.}(2016){Mann}, {Gaidos}, {Mace}, {Johnson}, {Bowler},
  {LaCourse}, {Jacobs}, {Vanderburg}, {Kraus}, {Kaplan}, \& {Jaffe}}]{Mann2016}
{Mann}, A.~W., {Gaidos}, E., {Mace}, G.~N., {et~al.} 2016,
  \href{http://dx.doi.org/10.3847/0004-637X/818/1/46}{\JournalTitle{\apj}, 818,
  46}

\bibitem[{{Markwardt}(2009)}]{Markwart2009}
{Markwardt}, C.~B. 2009, in Astronomical Society of the Pacific Conference
  Series, Vol. 411, Astronomical Data Analysis Software and Systems XVIII, ed.
  D.~A. {Bohlender}, D.~{Durand}, \& P.~{Dowler}, 251

\bibitem[{{Martin} {et~al.}(2016){Martin}, {Lubow}, {Nixon}, \&
  {Armitage}}]{2016MNRAS.tmp..399M}
{Martin}, R.~G., {Lubow}, S.~H., {Nixon}, C., \& {Armitage}, P.~J. 2016,
  \href{http://dx.doi.org/10.1093/mnras/stw605}{\JournalTitle{\mnras}},
  \href{http://arxiv.org/abs/1603.03135}{{\sffamily arXiv:1603.03135
  [astro-ph.EP]}}

\bibitem[{{Morton} \& {Swift}(2014)}]{Morton2014a}
{Morton}, T.~D., \& {Swift}, J. 2014,
  \href{http://dx.doi.org/10.1088/0004-637X/791/1/10}{\JournalTitle{\apj}, 791,
  10}

\bibitem[{{Morton} \& {Winn}(2014)}]{Morton2014b}
{Morton}, T.~D., \& {Winn}, J.~N. 2014,
  \href{http://dx.doi.org/10.1088/0004-637X/796/1/47}{\JournalTitle{\apj}, 796,
  47}

\bibitem[{{Mulders} {et~al.}(2015){Mulders}, {Pascucci}, \&
  {Apai}}]{Mulders2015}
{Mulders}, G.~D., {Pascucci}, I., \& {Apai}, D. 2015,
  \href{http://dx.doi.org/10.1088/0004-637X/798/2/112}{\JournalTitle{\apj},
  798, 112}

\bibitem[{{Nagasawa} {et~al.}(2008){Nagasawa}, {Ida}, \&
  {Bessho}}]{2008ApJ...678..498N}
{Nagasawa}, M., {Ida}, S., \& {Bessho}, T. 2008,
  \href{http://dx.doi.org/10.1086/529369}{\JournalTitle{\apj}, 678, 498}

\bibitem[{{Newton} {et~al.}(2016){Newton}, {Irwin}, {Charbonneau},
  {Berta-Thompson}, {Dittmann}, \& {West}}]{Newton2016}
{Newton}, E.~R., {Irwin}, J., {Charbonneau}, D., {et~al.} 2016,
  \href{http://dx.doi.org/10.3847/0004-637X/821/2/93}{\JournalTitle{\apj}, 821,
  93}

\bibitem[{{Nutzman} \& {Charbonneau}(2008)}]{Nutzman:2008gf}
{Nutzman}, P., \& {Charbonneau}, D. 2008,
  \href{http://dx.doi.org/10.1086/533420}{\JournalTitle{\pasp}, 120, 317}

\bibitem[{{Ogihara} {et~al.}(2015){Ogihara}, {Morbidelli}, \&
  {Guillot}}]{2015A&A...578A..36O}
{Ogihara}, M., {Morbidelli}, A., \& {Guillot}, T. 2015,
  \href{http://dx.doi.org/10.1051/0004-6361/201525884}{\JournalTitle{\aap},
  578, A36}

\bibitem[{{Park} {et~al.}(2014){Park}, {Jaffe}, {Yuk}, {Chun}, {Pak}, {Kim},
  {Pavel}, {Lee}, {Oh}, {Jeong}, {Sim}, {Lee}, {Nguyen Le}, {Strubhar},
  {Gully-Santiago}, {Oh}, {Cha}, {Moon}, {Park}, {Brooks}, {Ko}, {Han}, {Nah},
  {Hill}, {Lee}, {Barnes}, {Yu}, {Kaplan}, {Mace}, {Kim}, {Lee}, {Hwang}, \&
  {Park}}]{Park2014}
{Park}, C., {Jaffe}, D.~T., {Yuk}, I.-S., {et~al.} 2014,
  \href{http://dx.doi.org/10.1117/12.2056431}{in Society of Photo-Optical
  Instrumentation Engineers (SPIE) Conference Series, Vol. 9147, Society of
  Photo-Optical Instrumentation Engineers (SPIE) Conference Series}, 1

\bibitem[{{Parviainen} \& {Aigrain}(2015)}]{2015MNRAS.453.3821P}
{Parviainen}, H., \& {Aigrain}, S. 2015,
  \href{http://dx.doi.org/10.1093/mnras/stv1857}{\JournalTitle{\mnras}, 453,
  3821}

\bibitem[{{Paulson} {et~al.}(2004){Paulson}, {Cochran}, \&
  {Hatzes}}]{2004AJ....127.3579P}
{Paulson}, D.~B., {Cochran}, W.~D., \& {Hatzes}, A.~P. 2004,
  \href{http://dx.doi.org/10.1086/420710}{\JournalTitle{\aj}, 127, 3579}

\bibitem[{{Pecaut} {et~al.}(2012){Pecaut}, {Mamajek}, \&
  {Bubar}}]{2012ApJ...746..154P}
{Pecaut}, M.~J., {Mamajek}, E.~E., \& {Bubar}, E.~J. 2012,
  \href{http://dx.doi.org/10.1088/0004-637X/746/2/154}{\JournalTitle{\apj},
  746, 154}

\bibitem[{{Perryman} {et~al.}(2001){Perryman}, {de Boer}, {Gilmore}, {H{\o}g},
  {Lattanzi}, {Lindegren}, {Luri}, {Mignard}, {Pace}, \& {de
  Zeeuw}}]{2001A&A...369..339P}
{Perryman}, M.~A.~C., {de Boer}, K.~S., {Gilmore}, G., {et~al.} 2001,
  \href{http://dx.doi.org/10.1051/0004-6361:20010085}{\JournalTitle{\aap}, 369,
  339}

\bibitem[{{Preibisch} {et~al.}(2002){Preibisch}, {Brown}, {Bridges},
  {Guenther}, \& {Zinnecker}}]{2002AJ....124..404P}
{Preibisch}, T., {Brown}, A.~G.~A., {Bridges}, T., {Guenther}, E., \&
  {Zinnecker}, H. 2002,
  \href{http://dx.doi.org/10.1086/341174}{\JournalTitle{\aj}, 124, 404}

\bibitem[{{Preibisch} {et~al.}(2001){Preibisch}, {Guenther}, \&
  {Zinnecker}}]{2001AJ....121.1040P}
{Preibisch}, T., {Guenther}, E., \& {Zinnecker}, H. 2001,
  \href{http://dx.doi.org/10.1086/318774}{\JournalTitle{\aj}, 121, 1040}

\bibitem[{{Quinn} {et~al.}(2012{\natexlab{a}}){Quinn}, {Bakos}, {Hartman},
  {Torres}, {Kov{\'a}cs}, {Latham}, {Noyes}, {Fischer}, {Johnson}, {Marcy},
  {Howard}, {Szentgyorgyi}, {F{\H u}r{\'e}sz}, {Buchhave}, {B{\'e}ky},
  {Sasselov}, {Stefanik}, {Perumpilly}, {Everett}, {L{\'a}z{\'a}r}, {Papp}, \&
  {S{\'a}ri}}]{Quinn:2012mz}
{Quinn}, S.~N., {Bakos}, G.~{\'A}., {Hartman}, J., {et~al.} 2012{\natexlab{a}},
  \href{http://dx.doi.org/10.1088/0004-637X/745/1/80}{\JournalTitle{\apj}, 745,
  80}

\bibitem[{{Quinn} {et~al.}(2012{\natexlab{b}}){Quinn}, {White}, {Latham},
  {Buchhave}, {Cantrell}, {Dahm}, {F{\H u}r{\'e}sz}, {Szentgyorgyi}, {Geary},
  {Torres}, {Bieryla}, {Berlind}, {Calkins}, {Esquerdo}, \&
  {Stefanik}}]{2012ApJ...756L..33Q}
{Quinn}, S.~N., {White}, R.~J., {Latham}, D.~W., {et~al.} 2012{\natexlab{b}},
  \href{http://dx.doi.org/10.1088/2041-8205/756/2/L33}{\JournalTitle{\apjl},
  756, L33}

\bibitem[{{Quinn} {et~al.}(2014){Quinn}, {White}, {Latham}, {Buchhave},
  {Torres}, {Stefanik}, {Berlind}, {Bieryla}, {Calkins}, {Esquerdo}, {F{\H
  u}r{\'e}sz}, {Geary}, \& {Szentgyorgyi}}]{Quinn2014}
---. 2014,
  \href{http://dx.doi.org/10.1088/0004-637X/787/1/27}{\JournalTitle{\apj}, 787,
  27}

\bibitem[{{Quirrenbach} {et~al.}(2010){Quirrenbach}, {Amado}, {Mandel},
  {Caballero}, {Mundt}, {Ribas}, {Reiners}, {Abril}, {Aceituno}, {Afonso},
  {Barrado Y Navascues}, {Bean}, {B{\'e}jar}, {Becerril}, {B{\"o}hm},
  {C{\'a}rdenas}, {Claret}, {Colom{\'e}}, {Costillo}, {Dreizler},
  {Fern{\'a}ndez}, {Francisco}, {Galad{\'{\i}}}, {Garrido}, {Gonz{\'a}lez
  Hern{\'a}ndez}, {Gu{\`a}rdia}, {Guenther}, {Guti{\'e}rrez-Soto}, {Joergens},
  {Hatzes}, {Helmling}, {Henning}, {Herrero}, {K{\"u}rster}, {Laun}, {Lenzen},
  {Mall}, {Martin}, {Mart{\'{\i}}n-Ruiz}, {Mirabet}, {Montes}, {Morales},
  {Morales Mu{\~n}oz}, {Moya}, {Naranjo}, {Rabaza}, {Ram{\'o}n}, {Rebolo},
  {Reffert}, {Rodler}, {Rodr{\'{\i}}guez}, {Rodr{\'{\i}}guez Trinidad},
  {Rohloff}, {S{\'a}nchez Carrasco}, {Schmidt}, {Seifert}, {Setiawan},
  {Solano}, {Stahl}, {Storz}, {Su{\'a}rez}, {Thiele}, {Wagner}, {Wiedemann},
  {Zapatero Osorio}, {Del Burgo}, {S{\'a}nchez-Blanco}, \&
  {Xu}}]{2010SPIE.7735E..37Q}
{Quirrenbach}, A., {Amado}, P.~J., {Mandel}, H., {et~al.} 2010,
  \href{http://dx.doi.org/10.1117/12.857777}{in Society of Photo-Optical
  Instrumentation Engineers (SPIE) Conference Series, Vol. 7735, Society of
  Photo-Optical Instrumentation Engineers (SPIE) Conference Series}

\bibitem[{{Raymond} \& {Cossou}(2014)}]{2014MNRAS.440L..11R}
{Raymond}, S.~N., \& {Cossou}, C. 2014,
  \href{http://dx.doi.org/10.1093/mnrasl/slu011}{\JournalTitle{\mnras}, 440,
  L11}

\bibitem[{{Rizzuto} {et~al.}(2016){Rizzuto}, {Ireland}, {Dupuy}, \&
  {Kraus}}]{2016ApJ...817..164R}
{Rizzuto}, A.~C., {Ireland}, M.~J., {Dupuy}, T.~J., \& {Kraus}, A.~L. 2016,
  \href{http://dx.doi.org/10.3847/0004-637X/817/2/164}{\JournalTitle{\apj},
  817, 164}

\bibitem[{{Rizzuto} {et~al.}(2015){Rizzuto}, {Ireland}, \&
  {Kraus}}]{2015MNRAS.448.2737R}
{Rizzuto}, A.~C., {Ireland}, M.~J., \& {Kraus}, A.~L. 2015,
  \href{http://dx.doi.org/10.1093/mnras/stv207}{\JournalTitle{\mnras}, 448,
  2737}

\bibitem[{{Rizzuto} {et~al.}(2011){Rizzuto}, {Ireland}, \&
  {Robertson}}]{2011MNRAS.416.3108R}
{Rizzuto}, A.~C., {Ireland}, M.~J., \& {Robertson}, J.~G. 2011,
  \href{http://dx.doi.org/10.1111/j.1365-2966.2011.19256.x}{\JournalTitle{\mnras},
  416, 3108}

\bibitem[{{Rogers} {et~al.}(2011){Rogers}, {Bodenheimer}, {Lissauer}, \&
  {Seager}}]{Rogers:2011lr}
{Rogers}, L.~A., {Bodenheimer}, P., {Lissauer}, J.~J., \& {Seager}, S. 2011,
  \href{http://dx.doi.org/10.1088/0004-637X/738/1/59}{\JournalTitle{\apj}, 738,
  59}

\bibitem[{{Schlaufman} {et~al.}(2009){Schlaufman}, {Lin}, \&
  {Ida}}]{2009ApJ...691.1322S}
{Schlaufman}, K.~C., {Lin}, D.~N.~C., \& {Ida}, S. 2009,
  \href{http://dx.doi.org/10.1088/0004-637X/691/2/1322}{\JournalTitle{\apj},
  691, 1322}

\bibitem[{{Schlawin} {et~al.}(2014){Schlawin}, {Herter}, {Henderson}, {Wilson},
  {Probst}, {Sprayberry}, {Bonati}, {Schurter}, {James}, {Warner}, {Tighe},
  {Adams}, \& {Mart{\'{\i}}nez}}]{2014SPIE.9147E..2HS}
{Schlawin}, E., {Herter}, T.~L., {Henderson}, C., {et~al.} 2014,
  \href{http://dx.doi.org/10.1117/12.2055233}{in \procspie, Vol. 9147,
  Ground-based and Airborne Instrumentation for Astronomy V}, 91472H

\bibitem[{{Schlegel} {et~al.}(1998){Schlegel}, {Finkbeiner}, \&
  {Davis}}]{1998ApJ...500..525S}
{Schlegel}, D.~J., {Finkbeiner}, D.~P., \& {Davis}, M. 1998,
  \href{http://dx.doi.org/10.1086/305772}{\JournalTitle{\apj}, 500, 525}

\bibitem[{{Skrutskie} {et~al.}(2006){Skrutskie}, {Cutri}, {Stiening},
  {Weinberg}, {Schneider}, {Carpenter}, {Beichman}, {Capps}, {Chester},
  {Elias}, {Huchra}, {Liebert}, {Lonsdale}, {Monet}, {Price}, {Seitzer},
  {Jarrett}, {Kirkpatrick}, {Gizis}, {Howard}, {Evans}, {Fowler}, {Fullmer},
  {Hurt}, {Light}, {Kopan}, {Marsh}, {McCallon}, {Tam}, {Van Dyk}, \&
  {Wheelock}}]{Skrutskie2006}
{Skrutskie}, M.~F., {Cutri}, R.~M., {Stiening}, R., {et~al.} 2006,
  \href{http://dx.doi.org/10.1086/498708}{\JournalTitle{\aj}, 131, 1163}

\bibitem[{{Slesnick} {et~al.}(2006){Slesnick}, {Carpenter}, \&
  {Hillenbrand}}]{slesnik06}
{Slesnick}, C.~L., {Carpenter}, J.~M., \& {Hillenbrand}, L.~A. 2006,
  \href{http://dx.doi.org/10.1086/503560}{\JournalTitle{\aj}, 131, 3016}

\bibitem[{{Tanaka} \& {Ward}(2004)}]{2004ApJ...602..388T}
{Tanaka}, H., \& {Ward}, W.~R. 2004,
  \href{http://dx.doi.org/10.1086/380992}{\JournalTitle{\apj}, 602, 388}

\bibitem[{{Vacca} {et~al.}(2003){Vacca}, {Cushing}, \& {Rayner}}]{Vacca2003}
{Vacca}, W.~D., {Cushing}, M.~C., \& {Rayner}, J.~T. 2003,
  \href{http://dx.doi.org/10.1086/346193}{\JournalTitle{\pasp}, 115, 389}

\bibitem[{{Vacca} {et~al.}(2004){Vacca}, {Cushing}, \&
  {Rayner}}]{2004PASP..116..352V}
---. 2004, \href{http://dx.doi.org/10.1086/382906}{\JournalTitle{\pasp}, 116,
  352}

\bibitem[{{van Eyken} {et~al.}(2012){van Eyken}, {Ciardi}, {von Braun}, {Kane},
  {Plavchan}, {Bender}, {Brown}, {Crepp}, {Fulton}, {Howard}, {Howell},
  {Mahadevan}, {Marcy}, {Shporer}, {Szkody}, {Akeson}, {Beichman}, {Boden},
  {Gelino}, {Hoard}, {Ram{\'{\i}}rez}, {Rebull}, {Stauffer}, {Bloom}, {Cenko},
  {Kasliwal}, {Kulkarni}, {Law}, {Nugent}, {Ofek}, {Poznanski}, {Quimby},
  {Walters}, {Grillmair}, {Laher}, {Levitan}, {Sesar}, \&
  {Surace}}]{2012ApJ...755...42V}
{van Eyken}, J.~C., {Ciardi}, D.~R., {von Braun}, K., {et~al.} 2012,
  \href{http://dx.doi.org/10.1088/0004-637X/755/1/42}{\JournalTitle{\apj}, 755,
  42}

\bibitem[{{Vanderburg} \& {Johnson}(2014)}]{Vanderburg2014}
{Vanderburg}, A., \& {Johnson}, J.~A. 2014,
  \href{http://dx.doi.org/10.1086/678764}{\JournalTitle{\pasp}, 126, 948}

\bibitem[{{Vanderburg} {et~al.}(2016){Vanderburg}, {Latham}, {Buchhave},
  {Bieryla}, {Berlind}, {Calkins}, {Esquerdo}, {Welsh}, \&
  {Johnson}}]{2016ApJS..222...14V}
{Vanderburg}, A., {Latham}, D.~W., {Buchhave}, L.~A., {et~al.} 2016,
  \href{http://dx.doi.org/10.3847/0067-0049/222/1/14}{\JournalTitle{\apjs},
  222, 14}

\bibitem[{{Vanhollebeke} {et~al.}(2009){Vanhollebeke}, {Groenewegen}, \&
  {Girardi}}]{Vanhollebeke2009}
{Vanhollebeke}, E., {Groenewegen}, M.~A.~T., \& {Girardi}, L. 2009,
  \href{http://dx.doi.org/10.1051/0004-6361/20078472}{\JournalTitle{\aap}, 498,
  95}

\bibitem[{{Ward}(1997)}]{1997ApJ...482L.211W}
{Ward}, W.~R. 1997,
  \href{http://dx.doi.org/10.1086/310701}{\JournalTitle{\apjl}, 482, L211}

\bibitem[{{Wizinowich} {et~al.}(2000){Wizinowich}, {Acton}, {Shelton},
  {Stomski}, {Gathright}, {Ho}, {Lupton}, {Tsubota}, {Lai}, {Max}, {Brase},
  {An}, {Avicola}, {Olivier}, {Gavel}, {Macintosh}, {Ghez}, \&
  {Larkin}}]{2000PASP..112..315W}
{Wizinowich}, P., {Acton}, D.~S., {Shelton}, C., {et~al.} 2000,
  \href{http://dx.doi.org/10.1086/316543}{\JournalTitle{\pasp}, 112, 315}

\bibitem[{{Wright} {et~al.}(2010){Wright}, {Eisenhardt}, {Mainzer}, {Ressler},
  {Cutri}, {Jarrett}, {Kirkpatrick}, {Padgett}, {McMillan}, {Skrutskie},
  {Stanford}, {Cohen}, {Walker}, {Mather}, {Leisawitz}, {Gautier}, {McLean},
  {Benford}, {Lonsdale}, {Blain}, {Mendez}, {Irace}, {Duval}, {Liu}, {Royer},
  {Heinrichsen}, {Howard}, {Shannon}, {Kendall}, {Walsh}, {Larsen}, {Cardon},
  {Schick}, {Schwalm}, {Abid}, {Fabinsky}, {Naes}, \& {Tsai}}]{Wright:2010fk}
{Wright}, E.~L., {Eisenhardt}, P.~R.~M., {Mainzer}, A.~K., {et~al.} 2010,
  \href{http://dx.doi.org/10.1088/0004-6256/140/6/1868}{\JournalTitle{\aj},
  140, 1868}

\bibitem[{{Wu} \& {Murray}(2003)}]{2003ApJ...589..605W}
{Wu}, Y., \& {Murray}, N. 2003,
  \href{http://dx.doi.org/10.1086/374598}{\JournalTitle{\apj}, 589, 605}

\bibitem[{{Yelda} {et~al.}(2010){Yelda}, {Lu}, {Ghez}, {Clarkson}, {Anderson},
  {Do}, \& {Matthews}}]{Yelda2010}
{Yelda}, S., {Lu}, J.~R., {Ghez}, A.~M., {et~al.} 2010,
  \href{http://dx.doi.org/10.1088/0004-637X/725/1/331}{\JournalTitle{\apj},
  725, 331}

\bibitem[{{Yu} {et~al.}(2015){Yu}, {Winn}, {Gillon}, {Albrecht}, {Rappaport},
  {Bieryla}, {Dai}, {Delrez}, {Hillenbrand}, {Holman}, {Howard}, {Huang},
  {Isaacson}, {Jehin}, {Lendl}, {Montet}, {Muirhead}, {Sanchis-Ojeda}, \&
  {Triaud}}]{2015ApJ...812...48Y}
{Yu}, L., {Winn}, J.~N., {Gillon}, M., {et~al.} 2015,
  \href{http://dx.doi.org/10.1088/0004-637X/812/1/48}{\JournalTitle{\apj}, 812,
  48}

\bibitem[{{Zacharias} {et~al.}(2012){Zacharias}, {Finch}, {Girard}, {Henden},
  {Bartlett}, {Monet}, \& {Zacharias}}]{Zacharias:2012vn}
{Zacharias}, N., {Finch}, C.~T., {Girard}, T.~M., {et~al.} 2012,
  \JournalTitle{VizieR Online Data Catalog}, 1322, 0

\end{thebibliography}
